\documentclass[aps,secnumarabic,pra,twocolumn,amsmath,amssymb,nofootinbib,eqsecnum,superscriptaddress,nobibnotes,floatfix]{revtex4-1}
\usepackage{float}
\usepackage[squaren]{SIunits}
\usepackage{verbatim}
\usepackage{mathrsfs}
\usepackage{comment}
\usepackage{braket}
\usepackage{bbold}
\usepackage{amsmath}
\usepackage[colorlinks]{hyperref}
\usepackage{pdfpages}
\usepackage{xcolor}
\usepackage[latin1]{inputenc}
\usepackage{graphicx}
\usepackage{amssymb}
\usepackage{amsmath}
\usepackage{amsthm}
\usepackage{xspace}
\usepackage{times}
\usepackage{longtable}
\usepackage{ulem}   
\newfont{\tensy}{cmsy10}

\newcommand{\ma}[1]{\begin{align}{#1}\end{align}}

\makeatletter

\newcommand{\Rmnum}[1]{\expandafter\@slowromancap\romannumeral #1@}
\makeatother

\begin{document}

\title{Signatures of topological quantum phase transitions in driven and dissipative qubit-arrays}

\date{\today}

\author{Y. L. Dong}
\affiliation{College of Physics, Optoelectronics and Energy, Soochow University, Suzhou, Jiangsu 215006, P. R. China}

\author{Titus Neupert}
\affiliation{Princeton Center for Theoretical Science, Princeton University, Princeton, New Jersey 08544, USA}

\author{R. Chitra}
\affiliation{Institut f\"ur Theoretische Physik, ETH-Z\"urich, 8093 Z\"urich, Switzerland}

\author{Sebastian Schmidt}
\affiliation{Institut f\"ur Theoretische Physik, ETH-Z\"urich, 8093 Z\"urich, Switzerland}

\begin{abstract}

We study photonic signatures of symmetry broken and topological phases in a driven, dissipative circuit QED realization of spin-1/2 chains.
Specifically, we consider the  transverse-field XY model  and a dual model with 3-spin interactions. The former has a ferromagnetic and a paramagnetic phase, while the latter features, in addition, a symmetry protected topological phase. Using the method of third quantization, we calculate the non-equilibrium steady-state of the open spin chains for arbitrary system sizes and temperatures. We find that the bi-local correlation function  of the spins at both ends of the chain provides a sensitive measure for both symmetry-breaking and topological phase transitions of the systems, but no universal means to distinguish between the two types of transitions. 
Both models have equivalent representations in terms of free Majorana fermions, which host zero, one and two topological Majorana end modes in the paramagnetic, ferromagnetic, and symmetry protected topological phases, respectively. 
The correlation function we study retains its bi-local character in the fermionic representation, so that our results are equally applicable to the fermionic models in their own right. We propose a photonic realization of the dissipative transverse-field XY model in a tunable setup, where an array of superconducting transmon qubits is coupled at both ends to a photonic microwave circuit. 

\end{abstract} 


\maketitle

\section{Introduction}

The study of topological phase transitions and related phenomena has become a major focus of condensed matter
research in recent years \cite{Hasan2010,Qi2011}. In particular, Majorana bound states have attracted great attention due to potential
applications in topological quantum computing \cite{Kitaev2001,Nayak2008,Lutchyn2010}. However, the observation 
and coherent control of such exotic physics in traditional solid-state systems is difficult,
due to material imperfections, decoherence, and disorder \cite{Alicea2012}.  In contrast, quantum engineered 
 artificial systems  such as photonic crystals, coupled cavities and waveguides as well as artificial atoms provide a new and versatile 
platform for controlled quantum simulations of topological phenomena \cite{Lu2014}. 
Well known examples include the realization of unidirectional waveguides \cite{Wang2009}, the quantum spin Hall effect \cite{Hafezi2013}, Floquet topological insulators \cite{Rechtsman2013}, and photonic quasicrystals \cite{Vardeny2013,Tanese2014}.\\

Recently, a proposal to observe Majorana states in one dimensional  interacting cavity arrays was discussed in Ref.~\cite{Bardyn2012}.
Strong effective photonic interactions where shown to lead to a direct simulation of the transverse field Ising model.
In contrast  to related studies of quantum spin chains \cite{Levitov2001,Zvyagin2013} and light-matter systems in the ultra-strong coupling regime \cite{Hwang2013,Zhu2013}, the cavity array in Ref. \cite{Bardyn2012} is inherently driven and dissipative and thus settles in a non-equilibrium steady state. Dissipation in the cavities permits non-demolition measurements of various observables  via  the detection of photons emitted by the system. Numerical studies carried out for small system sizes suggest that the presence of Majorana-like modes manifests itself in a peak in the bi-local, second-order cross correlation function of the photons. This interesting result motivates the study of larger system sizes and related models with richer topological features in order to investigate how generic and robust this signature is for the study of topological phenomena, including Majorana bound states, in driven, dissipative systems.\\

In this paper, we  focus on two  related one dimensional  systems:  the  transverse-field XY model (TXY) and a model with three spin interactions (3SI)  with dissipation. The latter  is related to the TXY model by a duality transformation \cite{Niu2012}.  In the absence of dissipation, the ground state of both models features various quantum phase transitions and topologically non-trivial phases.  We study single spin observables as well as cross correlation functions in the nonequilibrium steady state to see if these are
capable of tracking all  ground state phase transitions, including those between phases with differing nontrivial topology.
We propose a tunable implementation of the TXY model based on an array of superconducting transmon qubits (compromising the system), which is coupled to external microwave transmission lines that serve as a dissipative bath. The array is pumped by a time-dependent magnetic flux and two external microwave drives acting on the boundary qubits. In the non-equilibrium steady state of such a driven, dissipative qubit chain, microwave photons are emitted into the transmission lines and can be detected using quantum-limited, parametric amplifiers \cite{Eichler2014}.\\

We use Lindblad master equations and the method of third quantization \cite{Prosen2008} to obtain the non-equilibrium steady state (NESS) of the dissipative chains for arbitrary system sizes.  We find that observables measured in the NESS  for weak dissipation help us to probe the underlying ground state
phase diagram of these models. In particular, the photon cross correlation function  provides a sensitive measure for the change of  ground state topology between different phases with differing topology as well as transitions between ordered magnetic phases with the same topology. However, 
 it does not allow us to distinguish between  
different phases far away from the  ground state phase boundaries. All phase transitions in these models are signaled by peaks  and/or oscillations in the photon cross correlator. The position and amplitude of these
features  depend sensitively on the size of the array as well as on temperature and the coupling parameters.

Below we give a brief outline of the paper. In Sec.~\ref{sec: model} we discuss the phases and topological properties of the TXY model and the 3SI model. For the latter, we provide a characterization in the language of  symmetry protected topological  phases. In Sec.~\ref{sec: experimental implementation}, we propose an in-situ tunable circuit QED implementation. Section~\ref{sec: third quantization} introduces the method of third quantization for the solution of the Lindblad-type master equation of the system's density matrix including the effects of drive and dissipation. Section~\ref{sec: spectroscopy} discusses a spectroscopic method for the detection of local and bi-local photonic correlation functions. In Sec.~\ref{sec: results}, we present the results of our numerical calculations and conclude our studies in Sec.~\ref{sec: conclusion}.\\

\section{Model}
\label{sec: model}
\begin{figure}[t]
\centering
\includegraphics[width=0.22\textwidth,clip]{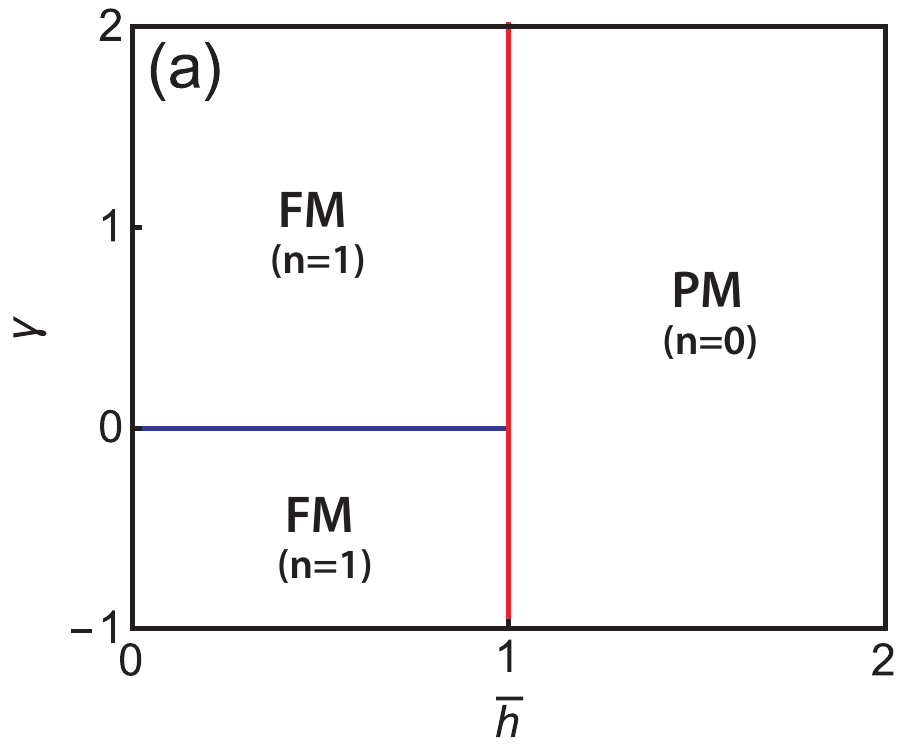}\hspace{0.1cm}
\includegraphics[width=0.22\textwidth,clip]{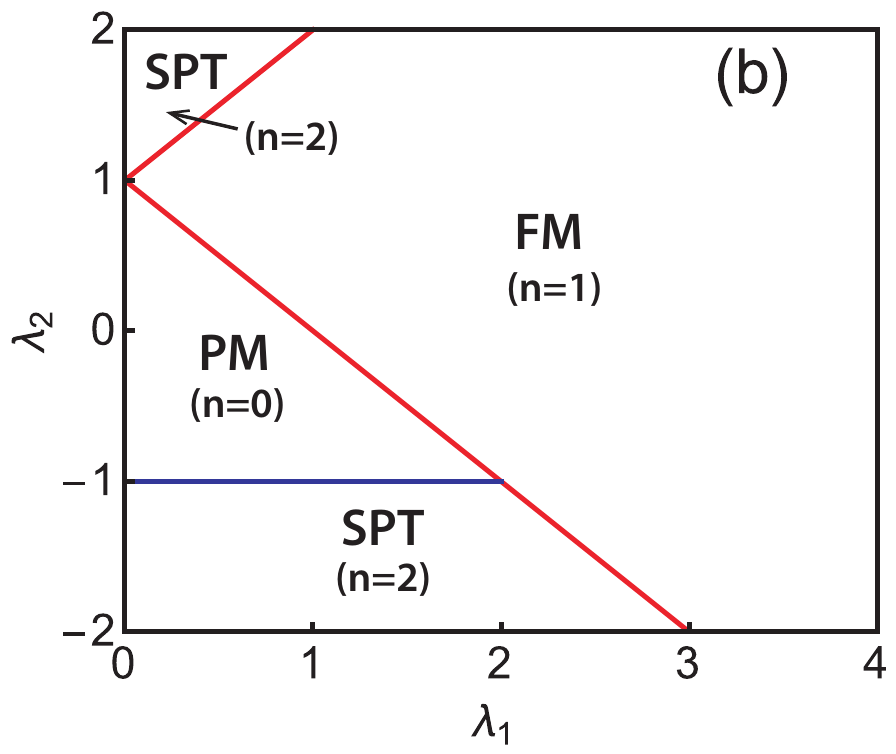}
\caption{(a) Phase diagram of the TXY model in Eq.~(\ref{model}) with anisotropy parameter $\gamma$ in a transverse magnetic field $\bar{h}$.
The red line marks the  transition between PM and FM phases. 
The PM and FM phases correspond to phases with $n=0$ and $n=1$ Majorana end modes in the fermionic counterpart, respectively.The blue line corresponds to the transition from a ferromagnet in  the x-direction to a ferromagnet in  the y-direction. (b) Phase diagram of the  3SI model in Eq.~(\ref{dual}) (see text below Eq.~(\ref{dual}) for definition parameter $\lambda_{1,2}$). The blue and red lines correspond to  the equivalent transitions  in (a), respectively. The 3SI model   also has a phase with  $n=2$ Majorana modes per end.}
\label{fig:pd}
\end{figure}

In this paper, we study physical signatures associated with phase transitions  in the anisotropic TXY model 
\begin{eqnarray}
H=\sum_i h_i \sigma^z_i + \sum_i J_{xi} \sigma^x_i \sigma^x_{i+1} + \sum_i J_{yi} \sigma^y_i \sigma^y_{i+1}\,
\label{model}
\end{eqnarray}
with $N$ spins described by the Pauli operators $\sigma^\alpha_i$ ($\alpha=x,y,z$) on lattice site $i=1,\cdots,N$, exchange coupling constants $J_{xi}$, $J_{yi}$ and transverse magnetic field $h_i$. For  the case of translationally invariant coupling constants, i.e., $J_{xi}=J_x$, $J_{yi}=J_y$, and $h_i=h$,  the well known  phase diagram is shown in Fig.~\ref{fig:pd} in terms of  the anisotropy parameter $\gamma=(J_x-J_y)/(J_x+J_y)$ and the rescaled transverse field $\bar{h}=h/(J_x+J_y)$.
The phase diagram features two quantum phase transitions, where the spectrum of the infinite chain becomes gapless: (i) the Ising transition at $\bar{h}=\pm 1$ between paramagnetic (PM) phase (for $|\bar{h}|>1$) and the ferromagnetic (FM) phase (for $|\bar{h}|<1$) and (ii) the anisotropic transition at $\gamma=0$, where the FM order flips between the x-direction and the y-direction. The two transition lines meet at a multi-critical point at $\gamma=0$ and $\bar{h}=1$ \cite{Dutta2015}.  For $\bar{h}<1$ and $\gamma\neq 0$, the chain has a two-fold degenerate, gapped ground state in the thermodynamic limit, while for $\bar{h} >1$ the ground state is non-degenerate. \\

This  model is an ideal example  to illustrate the difference between
Landau order and symmetry protected topological (SPT) phases. 
Crucial to this distinction is the notion of locality: 
since the  spin operators $\sigma^\pm$ and $\sigma_z$ are local, 
the ground state degeneracy seen for $\bar{h}<1$  is lifted by generic local perturbations, leading to a FM ground state by the mechanism of spontaneous symmetry breaking. The spin operators serve as a local observable to detect which ground state the system is in.\\

Using the non-local Jordan-Wigner transformation
\begin{eqnarray}
\label{jwtrans}
\sigma^-_i=c_ie^{i\pi\sum_{j<i} c_j^\dagger c_j}\quad\mbox{and}\quad \sigma^z_i=2c_i^\dagger c_i -1\,,
\end{eqnarray}
where the operators $c_i$ obey standard fermionic commutation relations,
Hamiltonian (2.1) can be rewritten as a  quadratic fermionic Hamiltonian
\begin{eqnarray}
H&=& 2 \sum_i h_i c_i^\dagger c_i  + \sum_i (J_{xi}+J_{yi}) (c_i^\dagger c_{i+1} +\text{h.c.})\nonumber\\
 &+& \sum_i (J_{xi}-J_{yi}) (c_i^\dagger c^\dagger_{i+1} + \text{h.c.})\,.
\label{fermionicmodel}
\end{eqnarray}
Though both Hamiltonians~\eqref{model} and~\eqref{fermionicmodel} share the same spectra, the transformation~\eqref{jwtrans} links two systems in which  either the fermion operators $c_i$  can be viewed as local and the spin operators $\sigma^\pm$  as nonlocal or vice versa. Thus, either Eq.~\eqref{model} or Eq.~\eqref{fermionicmodel} can be a local representation of the Hamiltonian in a given system. In particular, if the $c_i$ are the true local operators, the degenerate ground states for $\bar{h}<1$ discussed earlier cannot be distinguished by mere local measurements in the thermodynamic limit. In the fermionic representation, the ground state doublet is equivalent to a single fermionic level that can be either filled or empty.  The spectral weight of this level is exponentially localized at the \emph{two} ends of the chain rendering it manifestly nonlocal leading to  edge Majorana modes. The two ground states can be distinguished by their fermion parity. To render this notion more explicit, we transform the Hamiltonian~\eqref{fermionicmodel} to the Majorana basis 
\begin{eqnarray}
w_{2i-1}=c^\dagger_i+c_i\quad\mbox{and}\quad w_{2i}=-i(c_i^\dagger-c_i)\,,
\end{eqnarray}
where the Majorana operators satisfy the anti-commutation relations
\begin{equation}
\{w_{j},w_{k}\}=2\delta _{j,k}\qquad j,k=1,2,\ldots ,2n\,,
\end{equation}%
which yields for open boundary conditions 
\begin{eqnarray}
H&=&-i\sum_{i=1}^{N}h_{i}\omega _{2i-1}\omega _{2i}\\
&&-i\sum_{i=1}^{N-1}\left( J_{xi}\omega _{2i}\omega
_{2i+1}-J_{yi}\omega _{2i-1}\omega _{2i+2}\right)\nonumber\,.
\end{eqnarray}%
Solving the eigenvalue equation for the zero energy modes
\begin{eqnarray}
H\underline{w}=0\quad\mbox{with}\quad \underline{w}^T=(w_1,w_2,..,w_{2N})
\end{eqnarray}
gives a recursion relation for the components $w_i$, which can be easily solved
for the different phases of the Hamiltonian. In the FM phase one obtains two exponentially localized modes at the ends of the chain except for the anisotropy transition line with $\gamma=0$, where no  such Majorana mode exists (but the bulk is gapless).\\

We also study another interesting model, the 3SI model with longer-range three spin interactions \cite{Niu2012}  described by the Hamiltonian
 \begin{equation}
 \label{dual}
H_d=  J_x\sum_i \left( \sigma_i^z + \lambda_{1} \sigma_i^x \sigma_{i+1}^x +\lambda_{2} \sigma_{i-1}^x\sigma_{i}^z \sigma_{i+1}^x\right)\,.
\end{equation}
The 3SI model is in fact related to the TXY model via a duality transformation  provided the couplings read
$\lambda_1=h/J_{x}$ and $\lambda_2=-J_{y}/J_{x}$. 

Despite the three-spin interaction in Eq.~(\ref{dual}), this model can still be mapped to a quadratic fermionic Hamiltonian via the Jordan-Wigner transformation in Eq.~(\ref{jwtrans}), yielding a fermionic Hamiltonian with both nearest- and next-nearest neighbor hopping terms. The latter can be diagonalized and the resulting  phase diagram is depicted in Fig.\ref{fig:pd}(b).  The gap vanishes at the  three phase transition lines:  $\lambda_2=1\pm\lambda_1$
corresponding to the Ising transition at $\bar{h}=\pm 1$ and  $\lambda_2=-1$ for $\lambda_1<2$ which corresponds to the anisotropic transition at $\gamma=0$ and $\bar{h}<1$.
However, the duality transformation does not preserve the topological properties of the phases of the two models with open boundary conditions. In fact, the dual model features richer topological characteristics both in the spin and fermionic representations.

In the fermionic representation, the phase with small $|\lambda_1|$ and small $|\lambda_2|$ is topologically trivial with no Majorana end modes ($n=0$). The phase with large $|\lambda_1|$ is related to the Kitaev chain and hosts a single Majorana state at each end ($n=1$). The phase with large $|\lambda_2|$ can be thought of as two Kitaev chains, that are completely decoupled for $\lambda_1=0$, and consequently host $n=2$ Majorana states at each end. This is in line with the model belonging to symmetry class BDI in the classification of Refs.~\cite{Schnyder08} and~\cite{Ryu10}, due to an effective time-reversal symmetry. This symmetry class can support any integer number of Majorana modes as topologically stable end states.

In the spin basis, the $n=0$ and $n=1$ phases correspond to PM and FM phases, respectively (note that for $\lambda_2=0$ the Hamiltonian reduces to that of the transverse-field Ising model (TFIM)). 
The phase with large $|\lambda_2|$, in contrast, is in itself a topological phase of the spin model (a so-called symmetry protected topological, SPT, phase). The degeneracy arises from topologically protected end excitations of the spin chain and is not related to spontaneous symmetry breaking. Hamiltonian~\eqref{dual} has an effective time-reversal symmetry $T=\mathcal{K}\prod_{i=1}^N\sigma_z$, where $\mathcal{K}$ denotes complex conjugation. (The operator $\prod_{i=1}^N\sigma_z$ is another independent $\mathbb{Z}_2$ symmetry of the model, but this symmetry is not important for the topologically nontrivial phase we discuss.) The operator $T$ represents a $\mathbb{Z}_2^T$ symmetry in the classification of bosonic SPT phases of Ref.~\cite{Chen13}. In one dimension, this symmetry class has a $\mathbb{Z}_2$ topological classification, and the phase of Hamiltonian~\eqref{dual} with large $|\lambda_2|$ is a nontrivial example belonging to this class. To prove the existence of topological end excitations more explicitly, consider Hamiltonian~\eqref{dual} in the limit $J_x\to0$, $h=0$, so that only the term proportional to $\lambda_2$ is present. In this limit, the operators $\Sigma^x=\sigma^x_1$, $\Sigma^y=\sigma^y_1\sigma^x_2$, and $\Sigma^z=\sigma^z_1\sigma^x_2$ that are all localized at one end of the chain furnish a Pauli algebra that commutes with $H$.~\cite{Yao2015} (A similar set of local operators can be found on the other end of the chain.) At the same time, these operators are odd under $T$ and therefore not allowed as local perturbations to $H$ at the end of the chain. Consequently, in order to realize the Pauli algebra, the chain end has to host a two-fold degeneracy (a spin-1/2 end excitation) that is topologically protected. This end excitation carries a projective representation $T_{\mathrm{e}}$ of the time-reversal symmetry with $T_{\mathrm{e}}^2=-1$, while $T^2=+1$ in the bulk. Away from the special limit $J_x\to0$, $h=0$, the operators $\Sigma^x$, $\Sigma^y$, and $\Sigma^z$ will be dressed by exponentially decaying tails towards the bulk, but retain the symmetry properties outlined above. Thus, the phase with large $|\lambda_2|$ has the characteristic properties of a nontrivial SPT phase. 
In summary, the  3SI model in the spin representation hosts a topological phase, a trivial phase, and a symmetry broken phase.  

What kind of observables can one calculate that will be sensitive to symmetry breaking  transitions as well as topological transitions?
Based on the work of Ref.~\cite{Bardyn2012} we  
 study one-end and end-to-end correlation functions of this system in a non-equilibrium steady state when coupled to a bath. We study two observables which correspond to local/bilocal measurements in both spin and fermionic implementations of the model. Concretely, we are interested in the $z$-polarization of the end spin(s) which is equivalent to the occupation of a fermionic level at the end of the chain (in the fermionic representation). This polarization can either be measured on one end alone, or as an end-to-end correlation. We will show that the spin polarization at one end is sensitive to most of the phase transitions in our models. It captures the anisotropic phase transition in the TXY model and all phase transitions out of the paramagnetic phase in the 3SI model. It does, however, not allow to distinguish between the SPT and the FM phases of the 3SI model, for example. In contrast, the end-to-end spin correlations are enhanced at all phase transitions in the models we study, no matter whether they are of topological or symmetry-breaking nature. However, they cannot be used to discriminate the nature of the different phases, as they vanish far away from a phase boundary in all cases.

In the next section we will derive a realization of the  TXY model~\eqref{model} in a circuit QED setup using superconducting qubits.  Since the qubits directly realize spins,   the spin operators $\sigma^\alpha_i$ are local and the Jordan-Wigner fermions are non-local.

\section{Circuit QED implementation}
\label{sec: experimental implementation}

\begin{figure}[t]
\centering
\includegraphics[width=0.45\textwidth,clip]{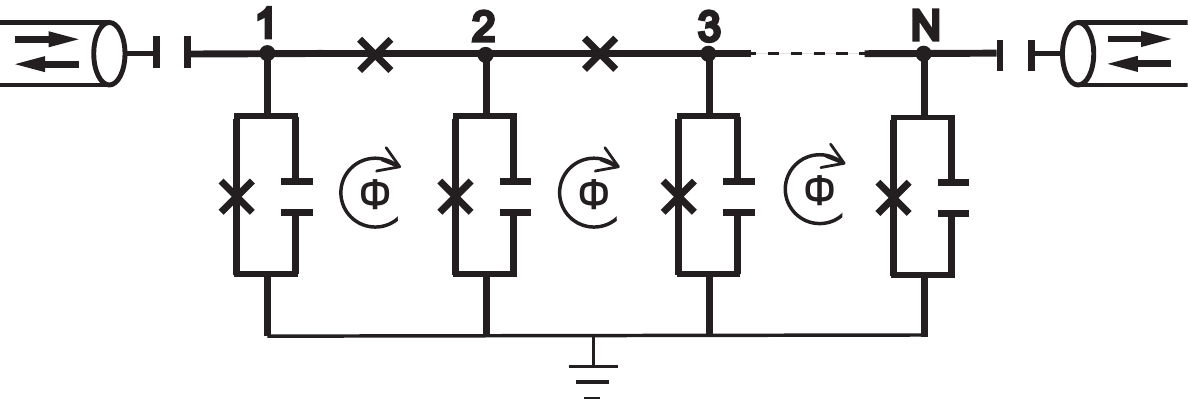}
\caption{Scheme of a 1D array of transmon qubits.  Each  transmon, comprising   a Josephson junction in parallel to a large capacitance, is grounded and connected to node $i$. Two neighbouring transmons at node $i$ and $i+1$ are coupled via a Josephson junction and enclose a common loop, which is threaded by an external magnetic flux $\Phi$ (we assume that the internal loop of the transmon remains unaffected by the external flux). The outer transmons at node $i=1,N$ are coupled capacitively to two external transmission lines, which act as a reservoir and an external drive.}
\label{fig:scheme}
\end{figure}

Coupling superconducting qubits to microwave circuitry yields a versatile architecture for 
universal quantum computation and the simulation of various spin models \cite{Salathe2015} as well as bosonic and fermionic Hubbard models \cite{Houck2012,Schmidt2013*2, Barends2015}.
Below we present an in-situ tunable coupling scheme for the realization of the 1D dissipative TXY model discussed above
based on Josephson junction coupled transmon qubits  \cite{Chen2014} as shown in Fig.~\ref{fig:scheme}. 
In this section, we will derive the model~(\ref{model}) following the general procedure of quantizing superconducting circuits~\cite{Devoret1995} and obtain the coupling constants in terms of the capacitances, Josephson energies and the applied magnetic flux depicted in Fig.~\ref{fig:scheme}. Dissipation is modelled by
two external transmission lines coupled to left and right end of the chain, respectively, allowing us
to directly probe  the full phase diagram shown in Fig.~\ref{fig:pd} via the measured statistics of microwave photons emitted into the transmission lines. 

The circuit Hamiltonian describing the qubit chain without external transmission lines can be written as $H=\sum_i H_i + \sum_i H_{i,i+1}$, where $H_i$ denotes the Hamiltonian of a single transmon associated with node $i$ in Fig.~\ref{fig:scheme}, while $H_{i,i+1}$ describes the coupling between two neighbouring nodes (qubits) mediated by a Josephson junction. Here, the Hamiltonian of a single transmon is given by \cite{Koch2007}
\begin{eqnarray}
\label{transmon}
H_i = \frac{\hat{Q}_i^2}{2 C_i} - E_{J,i}\cos(\hat{\phi}_i)\,,
\end{eqnarray}
where $\hat{n}_i$ measures the number of Cooper pairs that have tunneled across the Josephson junction of the transmon on site $i$ and $\hat{Q}_i = 2 e \hat{n}_i$ denotes the associated charge. Further, $C_i$ is the internal capacitance of the transmon and $E_{J,i}$ its Josephson energy.  In Eq.~(\ref{transmon}), $\hat{\phi}_i$ denotes the superconducting phase operator, which obeys the commutation relation $[\hat{\phi}_i, \hat{n}_i]=i$. Therefore, we can represent these variables in terms of bosonic creation and annihilation operators, e.g., $\hat{n}_i=i(a_i^\dagger - a_i)/(2\alpha_i)$ and $\hat{\phi}_i= \alpha_i (a_i^\dagger + a_i)$ with $[a_i,a_i^\dagger]=1$ and a normalization constant $\alpha_i=(2E_{C,i}/E_{J,i})^{1/4}$ [with the charging energy  $E_{C,i}=e^2/(2C_i)$]. In the so-called transmon regime with large internal capacitance and $E_{J,i}\gg E_{C,i}$ (i.e., $\alpha_i \ll 1$) one can expand the cosine potential in Eq.~(\ref{transmon}) for small angles yielding to leading order
\begin{eqnarray}
\label{transmon2}
H_i \approx \sqrt{8 E_{J,i} E_{C,i} } a_i^\dagger a_i - \frac{E_{C,i}}{12} (a_i + a_i^\dagger)^4\,,
\end{eqnarray}
where we have neglected an overall constant. The last term in (\ref{transmon2}) describes a small nonlinearity which makes the spectrum and eigenfunctions of the transmon weakly anharmonic. In the following, we will neglect higher transmon levels and replace the bosonic creation (annihilation) operators with raising (lowering) operators for a two-level system (TLS), i.e., $a_i\rightarrow \sigma_i^-$ and $a^\dagger_i\rightarrow \sigma_i^+$. Strictly speaking, such a (hard-core) approximation would be valid if the nonlinearity is infinite, i.e., $E_{C,i}\rightarrow\infty$, which formally contradicts our assumption $E_{J,i}\gg E_{C,i}$. However, we will excite the qubit array using coherent microwave tones with a well defined frequency and a drive amplitude which is smaller than the anharmonicity of the transmon. Thus, even a weak anharmonicity allows us (by a suitable choice of frequencies) to predominantly occupy only the two lowest levels of the transmon. In this case higher levels are only weakly excited and the two-level approximation gives meaningful results. We thus finally obtain the Hamiltonian of a transmon qubit \cite{Koch2007}
\begin{eqnarray}
\label{qubit}
H_i=\frac{\epsilon_i}{2} \sigma^z_i \quad \mbox{with} \quad \epsilon_i=\sqrt{8 E_{J,i} E_{C,i} }\,.
\end{eqnarray}

The coupling between two transmons is mediated by a Josephson junction and an external flux $\Phi$, which threads the common loop formed by the two qubits and the junction (see Fig.~\ref{fig:scheme}). The coupling Hamiltonian is given by 
\begin{eqnarray}
H_{i,i+1}= -E_{J,c}\cos{\left(\hat{\phi}_{i+1} - \hat{\phi}_{i} + \Phi_i\right)}\,,
\end{eqnarray}
where $E_{J,c}$ denotes the Josephson energies of the coupler junction and the external flux $\Phi_i$ is given in units of  $\Phi_0/(2\pi)=\hbar/(2e)$, where $\Phi_0$ denotes the magnetic flux quantum. By using the trigonometric
relation $\cos{\left(\phi_1- \phi_2\right)}=\cos{\left(\phi_1\right)}\cos{\left(\phi_2\right)}+\sin{\left(\phi_1\right)}\sin{\left(\phi_2\right)}$ and expanding the cosine/sine functions in small angles we obtain in a two-level approximation
\begin{eqnarray}
\sin{(\hat{\phi}_i)} \approx \alpha_i \sigma_i^x\,,\quad\cos{(\hat{\phi}_i)} \approx1-\alpha_i^2-\left(\alpha_i^2/2\right) \sigma^z_i.\nonumber\\
\end{eqnarray}
The direct coupling thus yields, to quadratic order in the $\alpha_i$, XX type interactions between neighboring qubits.
In order to activate couplings that describe the  $XY$ model, we will now consider a time-dependent external flux,
which is periodically modulated at the frequency difference and frequency sum of the two neighbouring qubits \cite{Kapit2015}, i.e.,
\begin{eqnarray}
\label{flux}
\Phi_i(t)=\frac{\pi}{2}-f_\Delta \cos{\left(\omega_i^- t\right)} - f_\Sigma \cos{\left(\omega_i^+ t\right)}
\end{eqnarray}
with frequencies $\omega_i^\pm=\epsilon_i \pm \epsilon_{i+1}$ and where $\epsilon_i$ denotes the eigenfrequency of each transmon defined in Eq.~(\ref{qubit}).
We assume a weak drive with $f_{\Delta}, f_{\Sigma} \ll 1$ and expand the cosine of the flux to first order in $f_{\Delta}, f_{\Sigma}$.
In the interaction picture with respect to $H_i$ the total Hamiltonian has then time-independent as well as time-dependent terms.
By neglecting the fast rotating terms it is straightforward to show that the time-averaged effective Hamiltonian is given by Eq.~(\ref{model})
with the parameters
\begin{eqnarray}
h_i=\epsilon_i/2\,,\quad J_{x(y),i}=E_{J,c} \alpha_i \alpha_{i+1}\left(f_\Delta \pm f_\Sigma \right)/4\,,
\end{eqnarray}
where the couplings $J_{x(y),i}$ can be tuned in-situ and locally (and we have neglected a small dispersive shift of the on-site frequency due to the interaction term). Also the qubit frequency $\epsilon_i$ can be tuned independently if the internal Josephson junction of the transmon is replaced by a SQUID loop threaded by an additional magnetic field~\cite{Koch2007}.\\

The transmission lines in Fig.~\ref{fig:scheme} can be considered as two large reservoirs for microwave photons. At one end they couple to the
qubit chain such that microwave photons can be emitted from the outer qubits into the transmission line.  At the other end they are connected to the amplification chain, where weak microwave signals are converted into an electronic signal and can thus be detected.
The coupling of the transmission lines to the chain of qubits can be described by a Lindblad-type master equation
\begin{eqnarray}
\label{master}
\frac{\mathrm{d}\rho }{\mathrm{d}t}= -i \left[H, \rho\right] +\sum_{i=1,N}\left[\Gamma^\downarrow_i \mathcal{D}[\sigma^-_i]\rho + \Gamma^\uparrow_i \mathcal{D}[\sigma^+_i]\rho\right]
\end{eqnarray}
with the Lindblad operator
\begin{eqnarray}
\mathcal{D}[\sigma^\pm_i]\rho=
 2  \sigma^\pm_i \rho \sigma^\mp_i - \sigma^\pm_i\sigma^\mp_i \rho - \rho \sigma^\pm_i\sigma^\mp_i 
\end{eqnarray}
and decay rates $\Gamma^\downarrow_i=\Gamma_i \left( n_i + 1 \right)$ and $\Gamma^\uparrow_i=\Gamma_i n_i$, where $n_i$ denotes the thermal occupation of the bath (transmission line). The decay rates $\Gamma_i$ can in principle be expressed in terms of the impedance and the coupling capacitances between the boundary qubits and the transmission line. However, in the weak coupling limit they can be measured using straightforward transmission spectroscopy and are thus considered here as 
an input parameter of the device.

The statistics of the emitted microwave photons can then be measured in the amplification chain with standard circuit QED technology.
The results of such a measurement relate to the steady state expectation values of the internal qubit degrees of freedom via standard input/output theory \cite{Gardiner1985}.  For example, the intensity of the radiation emitted into the left transmission line in Fig.~\ref{fig:scheme} is directly proportional to the expectation value of the spin projection in $z$-direction of the left most qubit with $i=1$. Using interferometric techniques it is also possible to measure non-local correlations, e.g., spin-spin correlation functions (see Sec.~\ref{sec: results}).\\

In principle, the spins are also subject to other sources of dissipation, e.g., spontaneous emission and dephasing. However, in transmon qubits the decay rates associated with these  processes can be strongly suppressed with respect to the coupling to the external transmission lines \cite{Koch2007}.  This justifies  our assumption   that these dissipative channels can be neglected and that the nonequilibrium steady state (NESS) of the system is dominated by  our engineered  dissipation.\\

In this section, we have shown how to realize the TXY model using standard circuit QED technology. A similar realization for the 3SI model should be feasible, but requires the implementation of multi-qubit interactions \cite{Mezzacapo2014} and more complex connectivity between qubits \cite{Lechner2015}. A precise discussion of the coupling design and estimates of realistic experimental parameters is thus more involved and left for future work.

\section{Method of third quantization}
\label{sec: third quantization}

The master equation (\ref{master}) for the density matrix  of the   TXY  and 3SI models    with dissipation from the boundaries is given by
\begin{equation}
\frac{\mathrm{d}\rho }{\mathrm{d}t}={\mathcal{L}}\rho :=-i%
[H,\rho ]+\sum_{\mu=1,N }\left( 2L_{\mu }\rho L_{\mu }^{\dagger }-\{L_{\mu
}^{\dagger }L_{\mu },\rho \}\right).  
\label{eq:lind}
\end{equation}%
In the Majorana representation, the Hamiltonians for both models take the simple 
 quadratic  form \begin{eqnarray}
H &=&\sum_{j,k=1}^{2N}w_{j}H_{jk}w_{k}
={\underline{w}}^{\mathsf{T}}\cdot {\mathbf{H}}\,\cdot{%
\underline{w}} ,
\label{eq:hamil} 
\end{eqnarray}
where the $2N\times 2N$ matrix ${\mathbf{H}}$ is hermitian and purely imaginary, implying 
 ${\mathbf{H}}$ is antisymmetric ${\mathbf{H}}^{\mathsf{T}}=-{\mathbf{H}}$.
The bath operators $L_{\mu }$ for the end spins with $\mu=1,N$ are linear  in the Majorana operators 
\begin{eqnarray}
L_{\mu } &=&\sum_{j=1}^{2N}l_{\mu ,j}w_{j}=\underline{w}^{\mathsf{T}}\cdot \underline{l}_{\mu } 
\label{eq:lindb}
\end{eqnarray}%
with $\underline{l}_{\mu }$ being a complex vector of length $2N$. Note that this linear representation for the dissipator only holds for the end spins.
For spins in the chain, the dissipator comprises  of strings of Majorana operators  arising from the  non-local
nature of the Jordan-Wigner transformation. Neglecting the dissipation from the inner qubits, as discussed in the previous section, has the technical advantage of leading to quadratic
forms of the Liouvillian, which allows us to determine the NESS for arbitrary system size as will be explained below \cite{Prosen2008}.\\

The   quadratic Liouvillian problem of Eq.~(\ref{eq:lind}) can be reformulated in terms  the correlation matrix 
\begin{equation}
\label{correlation}
C_{i,j}=\frac{i}{2}\mathrm{tr}\left( \left[ \omega _{i},\omega _{j}\right],
\rho \right) .
\end{equation}%
which is 
a \emph{real anti-symmetric} matrix $\mathbf{C}\in \mathbb{R}^{2N\times 2N}$.  
The matrix $\mathbf{C}$ which determines all observables can be shown to obey  the Lyapunov equation \cite{Zunkovic2010}
\begin{equation}
\label{lyapunov}
\frac{d\mathbf{C}}{dt}=\mathbf{XC}+\mathbf{CX}^{\mathsf{T}}-\mathbf{Y},
\end{equation}%
where%
\begin{equation}
\begin{split}
\label{xymatrices}
\mathbf{X}&= -2(2i\mathbf{H}+\mathbf{M}+\mathbf{M}^{\mathsf{T}}),\\ 
 \mathbf{Y} &=4i\left( \mathbf{M}-\mathbf{M}^{\mathsf{T}}\right),\\
{\mathbf{M}} &=\sum_{\mu =1,N}\underline{l}_{\mu }\otimes \underline{l}_{\mu
}^{\dagger }\,.
\end{split}
\end{equation}%
It is straightforward to show that $\mathbf{X}$ and $\mathbf{Y}$ are real matrices
and ${\mathbf{M}}$ is a
complex Hermitian matrix.
For the anisotropic XY model with dissipation for the first and last spin, the only nonzero elements for $\mathbf{M}$ are%
\begin{eqnarray}
M_{1,1} =M_{2,2}=\Gamma _{1}^{+}/4,\quad M_{1,2} =M_{2,1}^{\ast }=-i\Gamma _{1}^{-}/4,
\end{eqnarray}
and
\begin{eqnarray}
M_{2N-1,2N-1} &=&M_{2N,2N}=\Gamma _{N}^{+}/4, \\
M_{2N-1,2N} &=&M_{2N,2N-1}^{\ast }=-i\Gamma _{N}^{-}/4,
\end{eqnarray}%
where
\begin{equation}
\Gamma _{1(N)}^{\pm}=\Gamma _{1(N)}^{\uparrow}\pm \Gamma _{1(N)}^{\downarrow}.\\
\end{equation}%
The stationary solution $\mathrm{d}\mathbf{C}/\mathrm{d}t=0$ fully
determines the NESS of the system.  All observables in the NESS are then easily obtained.
For example, the $z$-projection of the spin operators is given in terms of a product of Majorana operators and we can thus express its steady state value as
\begin{eqnarray}
\left\langle \sigma _{i}^{z}\right\rangle =-i\left\langle \omega
_{2i-1}\omega _{2i}\right\rangle= - C_{2i-1,2i}\,.
\label{sigma_z}
\end{eqnarray}
Expectation values of other observables can be obtained by applying Wicks theorem.

\section{Auxiliary qubit spectroscopy}
\label{sec: spectroscopy}

In general, the NESS of a driven, dissipative system corresponds to a mixture of several eigenstates of the Hamiltonian, even at zero bath temperature. However,
specific parts of the  energy spectrum  can be  probed via the measurement of  photons  emitted from the spectral region of interest by using filtering techniques. For example, in Ref.~\cite{Bardyn2012}, coupling to auxiliary cavities was proposed  as a way to filter photons that are predominantly emitted by the spectral region in which the lowest two Majorana states  reside.

This filtering scheme can be implemented by  engineering the end spins with $i=1,N$ so that they are subject to nearly zero transverse field and 
 weakly couple to the bulk spins $i=2,N-1$  through simple exchange interactions with
$J_{x,1},J_{x,N-1}\ll J_x$ and $J_{y,1},J_{y,N-1} \ll J_y$, while the 
 bulk spins are described by a homogeneous TXY model   with
\begin{equation}
\begin{split}
&h_{i}=h,\qquad i=2,\ldots, N-1,
\\
&J_{x,i}= J_x,\quad 
J_{y,i}= J_y,\qquad  i = 2,\ldots, N-2 . 
\end{split}
\end{equation}
 In this weak coupling scenario, the NESS is an admixture of the two  nearly degenerate lowest eigenstates of the bulk, if the bulk is in a gapped FM or SPT phase. The photon emission is dominated by transitions between these two lowest  states.

This spectroscopy scheme also works for nonzero transverse fields at the end cavities $h_{\mu},\ \mu=1,N,$  provided they are smaller than the finite-size splitting $\epsilon_\kappa$ between the quasi degenerate states in the FM or SPT  phase (see Appendix~\ref{app: exact solution}). However, this condition is hard to achieve by merely tuning charging and Josephson energies in Eq.~(\ref{qubit}). To circumvent this, we apply an additional drive  to the auxiliary qubits,  leading to the Hamiltonian $H_{\mu}=h_{\mu} \sigma^z_{\mu} + f ( \sigma^+_{\mu} e^{i \omega_d t} + {\rm h.c.})$, $\mu=1,N$, where $\omega_d$ is the drive frequency and $f$ the drive strength.
Using the unitary transformation $U_\mu=\exp{[i\omega_d \sigma^+_{\mu}\sigma^-_{\mu} t]}$, we obtain the auxiliary qubit Hamiltonian in the
rotating frame:
$H_{\mu}=\Delta_{\mu} \sigma^z_{\mu} + f \sigma^x_{\mu}$, with the shifted on-site frequency $\Delta_{\mu}=h_{\mu}-\omega_d$, for $\mu=1,N$. 
For drive amplitudes that obey $f \ll \Delta_{\mu} \ll \epsilon_\kappa$ the longitudinal field term can be neglected 
while simultaneously satisfying the requirement of small transverse fields.  We mention that in the experimental setup discussed in Sec.~\ref{sec: experimental implementation}, for the effective Hamiltonian to remain time-independent in the rotating frame, the frequency components of the time-modulated flux in Eq.~(\ref{flux}) for  the first and last SQUID  should be replaced by  $\omega_{1}^\pm=\Delta_{1} \pm \epsilon_{2}$ and $\omega_{N-1}^\pm=\epsilon_{N-1} \pm \Delta_{N}$.\\

In the setup with auxiliary qubits/spins, the statistics of the photons emitted from the boundaries of the chain can then be measured.  The first order moment of the photon distribution is the  intensity of  emitted photons, which is proportional to the $z$-component of the auxiliary spins. Using the third quantization formalism,  for the left auxiliary  qubit Eq.~(\ref{sigma_z}) yields
\begin{eqnarray}
\left\langle \sigma _{1}^{z}\right\rangle = - C_{1,2}\,.
\end{eqnarray}
The second order moment  is  the photon cross correlation
 function of the auxiliary spins, which involves nonlocal end-to-end spin correlations. Using third quantization and Wicks theorem, we
 find that the photon cross correlator is given by
\begin{equation}
\begin{split}
g_{1N}^{(2)}&\equiv1+\frac{\left\langle \sigma _{1}^{z}\sigma _{N}^{z}\right\rangle
-\left\langle \sigma _{1}^{z}\right\rangle \left\langle \sigma
_{N}^{z}\right\rangle}{\left( 1+\left\langle \sigma
_{1}^{z}\right\rangle \right) \left( 1+\left\langle \sigma
_{N}^{z}\right\rangle \right) }
\\
 &=1-\frac{C_{1,2N-1}C_{2,2N}-C_{1,2N}%
C_{2,2N-1}}{\left( 1-C_{1,2}\right) \left( 1-C%
_{2N-1,2N}\right) }\,.
\label{def:g2}
\end{split}
\end{equation}

In the following sections,  we  present numerical as well as analytical results for these two quantities for both the TXY and the 3SI models.  We will show that these quantities are sensitive to all the phase boundaries including boundaries
separating two topologically non-trivial phases.

\section{Results}
\label{sec: results}
\subsection{Transverse-field XY model}
\label{sec: XY model}

\begin{figure}[t]
\begin{center}
\includegraphics[width=0.23\textwidth,clip]{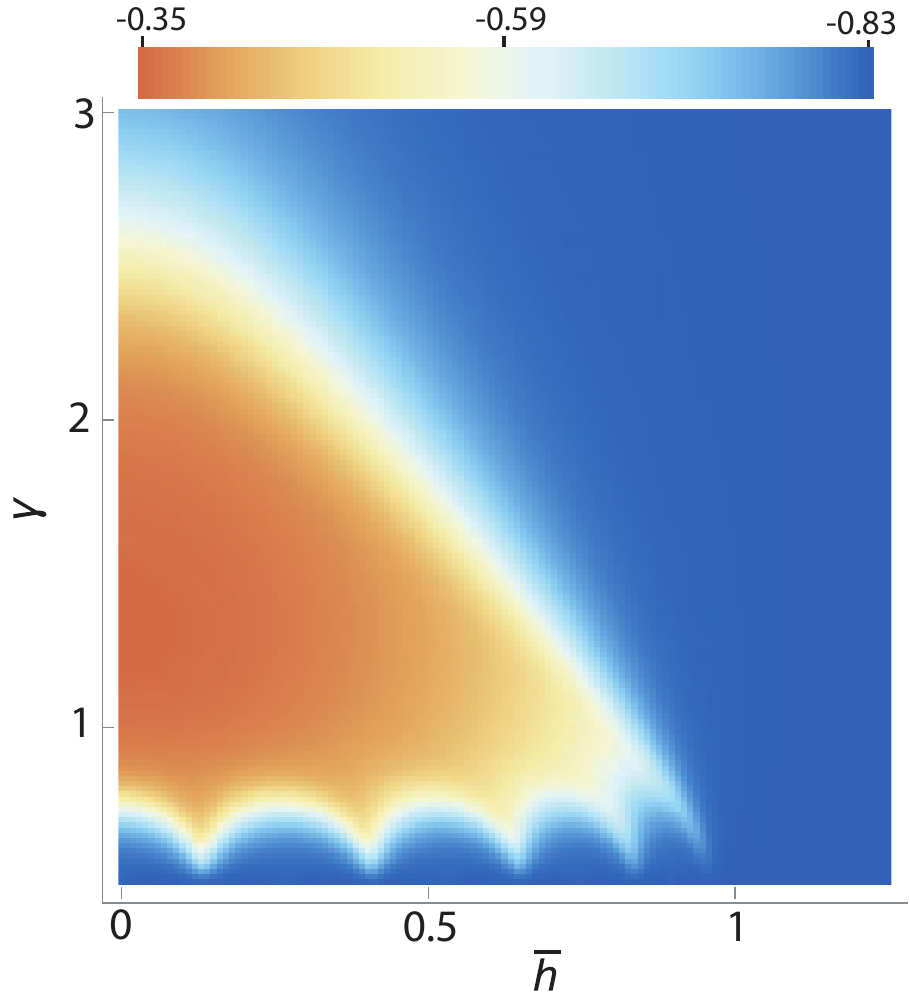}
\includegraphics[width=0.23\textwidth,clip]{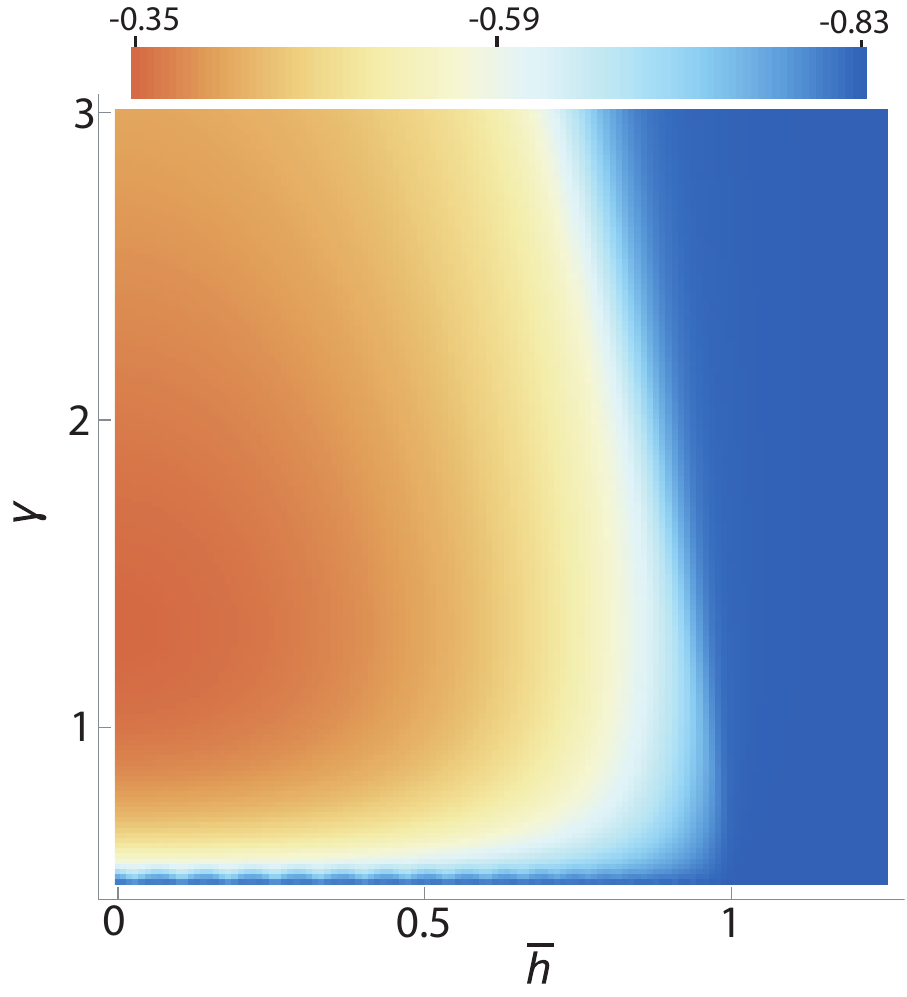}
\caption{
Auxiliary qubit occupation $\langle \sigma^z_1 \rangle$ (colour scale) as a function of bulk transverse field $\bar{h}$ and anisotropy parameter $\gamma=(J_x-J_y)/(J_x+J_y)$ for the NESS of the driven dissipative TXY model
with $N=12$ (left panel) and $N=42$ (right panel) spins, i.e., $N-2$ bulk spins and two auxiliary spins. The auxiliary spins are coupled weakly to the ends of the chain with $J_{\mu,1}=J_{\mu,N-1}=0.02 J_x$ ($\mu=x,y$) and even weaker to the external transmission lines (see Fig.~\ref{fig:scheme}) with dissipation rates $\Gamma_{1,N}=\Gamma=0.02 J_x$.
The number of thermal photons inside the transmission lines is assumed to be $n_{\rm th}=0.1$. Here, we have set $J_x=1$ such that a horizontal line with $\gamma=1$ corresponds to the special case of the transverse field Ising model with $J_y=0$. 
}
\label{fig:sz}
\end{center}
\end{figure}

\begin{figure}[t]
\begin{center}
\includegraphics[width=0.23\textwidth,clip]{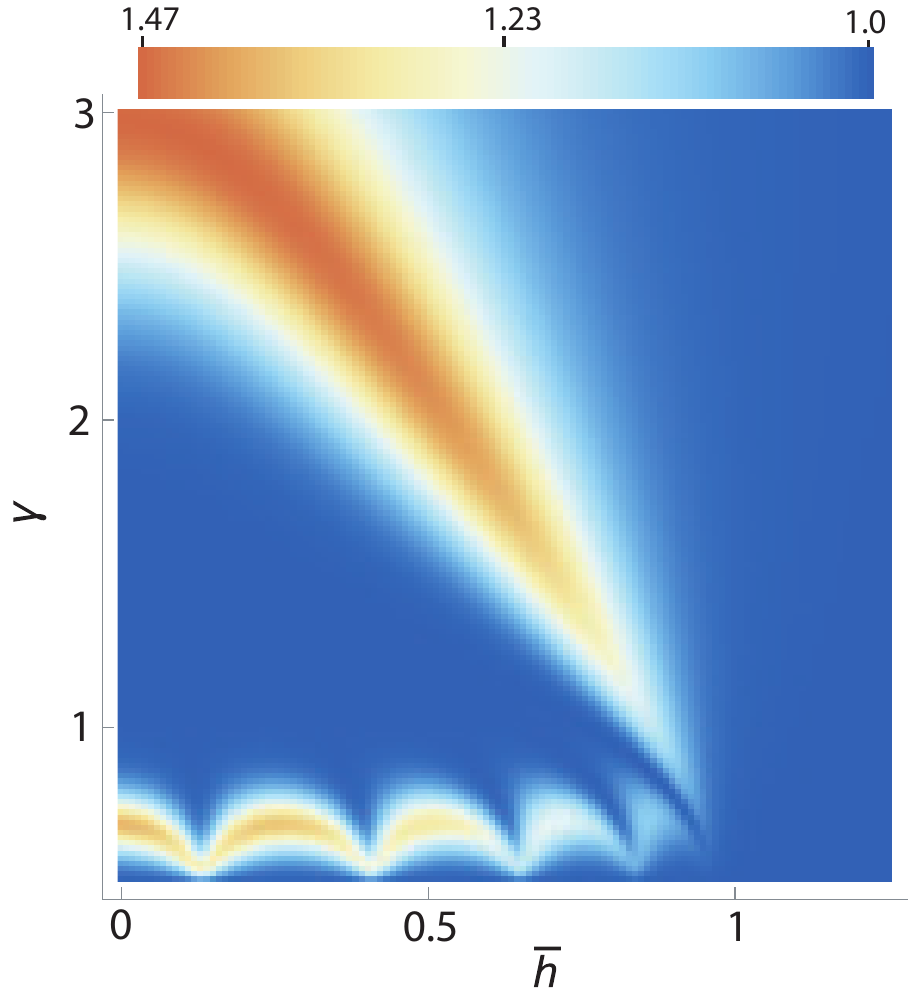}
\includegraphics[width=0.23\textwidth,clip]{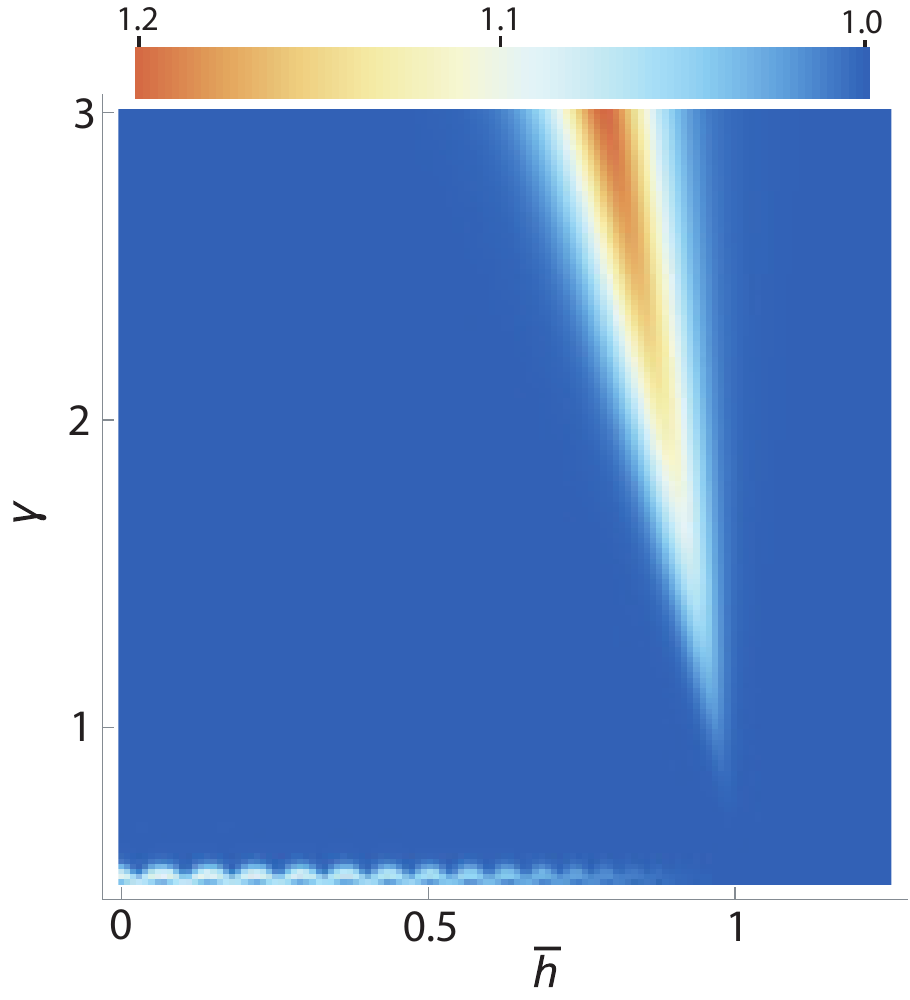}
\caption{
Second-order cross correlation function $g^{(2)}_{1N}$ (color scale) as defined in Eq.~(\ref{def:g2}) for $N=12$ (left panel) and $N=42$ (right panel). The same notation as in Fig.~\ref{fig:sz} applies.
}
\label{fig:g2}
\end{center}
\end{figure}

Our results for  the auxiliary qubit occupation and the end-to-end correlation function in the NESS of the TXY model  
are presented in Fig.~\ref{fig:sz} and Fig.~\ref{fig:g2} for the two different system sizes $N=12$ and $N=42$, and at a temperature low compared to the drive frequency such that the mean thermal photon number in the transmission lines is smaller than one, i.e., $n_{1},n_{N}\ll 1$. We first note that both quantities effectively trace the ground state phase diagram of the TXY model 
 without auxiliary qubits shown in Fig.~\ref{fig:pd}.
 The FM phase with $\bar{h}<1$ is associated with an almost saturated auxiliary spin with $\langle\sigma^z_{1,N}\rangle\approx 0$. In the PM phase with $\bar{h}>1$, the auxiliary spin resides in its ground-state at low enough temperatures $\langle\sigma^z_{1,N}\rangle\approx -1$. Finite-size effects tend to increase the region where $\langle\sigma^z_{1,N}\rangle\approx -1$, while in the scaling limit $N\rightarrow \infty$ there is a sharp transition between the two phases along the line $\bar{h}=1$. We also observe oscillations of both observables near the anisotropic transition line with a period that depends on the size of the chain. The photon cross correlation function does not allow to distinguish between the FM and PM phases away from the phase boundaries. However, it provides a sensitive measure for all phase transitions of the model as it shows a peak near the PM transition line and oscillations near the anisotropic line. Peak height and position, as well as oscillation period thereby sensitively depend on the size of the chain and the temperature of the bath as we  shall  show below.

The oscillations seen close to the phase transition boundary given by the $\gamma=0$ line for
${\bar h} <1$  arise from an interplay between incommensuration of the spin ordering and finite
size effects. In the absence of dissipation, as the $\gamma=0$ line is approached (both from
above and below), the gap in the dispersion $\epsilon_k$ (see Eq.~(6) in Ref.~\cite{Niu2012}) of the infinite chain with periodic boundary conditions closes at $k_0 a\equiv {\rm arccos}( - {\bar h } )$ indicating the existence of a soft mode or equivalently, low lying excitations, at $k_0$.  The momentum of the soft mode thus migrates from $k_0 a = 0$ to  $k_0 a = \pi$ as ${\bar h}$ takes values from $0$ to $1$ along the anisotropic phase boundary. This signals that the ground state of such a chain is a spin density wave with ordering wave vector $k_0$ exhibiting quasi long range order along this line.

\begin{figure}[t]                   
\centering
\includegraphics[width=0.3\textwidth,clip]{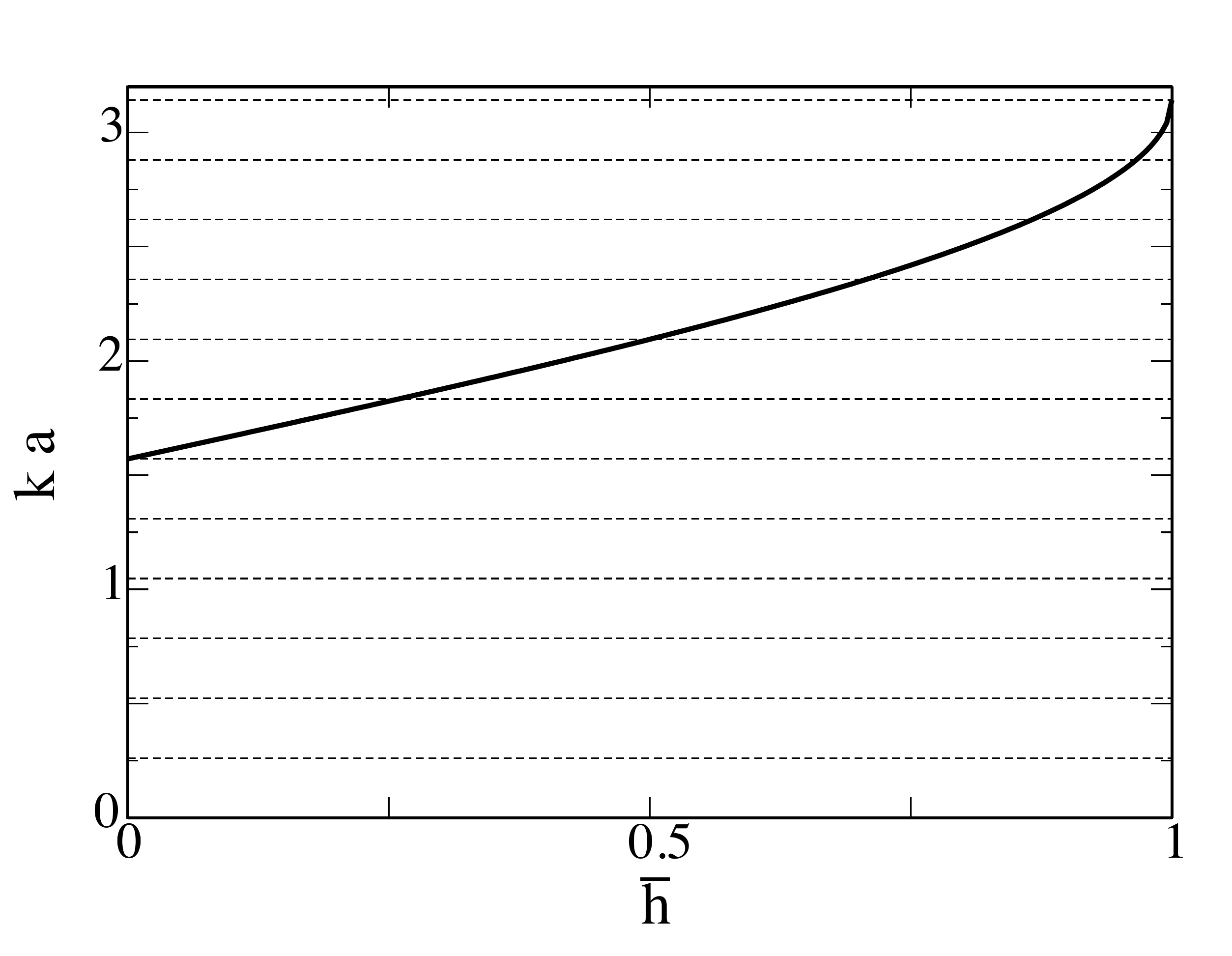}
\caption{\label{fig:k0}  Plot of ordering momentum $k_0$ and the set of discrete momenta $k_n$ for a $N=12$ TXY model. The values of $\bar{h}$, where $k_0$ and $k_n$ coincide correspond to the crests of the oscillations  near the anisotropy line in Figs.~\ref{fig:sz},\ref{fig:g2}.}
\end{figure}

For a finite size system and a generic value of $\bar{h}$, we have a nonzero value of the gap at $k_0$ due to the discretization of the momenta, as we now explain. The momenta are discrete and take the values  $ k_n =\pi n/aN$, where $N$ is the number of sites, $a$ is the lattice spacing, and the integer $n$ obeys $-N< n  \le N$. Depending on the system size and the value of $\bar{h}$, $k_0$  may or may not be included in the set of allowed discrete momenta $k_n$ and this has direct repercussions for expectation values of observables. It results in an oscillatory behavior of both  $\langle\sigma^z_{1,N}\rangle$ and $g^{(2)}_{1N}$ as seen in Figs.~\ref{fig:sz} and~\ref{fig:g2}.
When $k_0$ equals one of the $k_n$, the contribution of the soft mode leads to augmented values of both $\langle\sigma^z_{1,N}\rangle$ and $g^{(2)}_{1N}$ due to the existence of low-energy excitations, corresponding to the crests of the oscillations. The troughs correspond to cases where $k_0$ is most distant from any $k_n$ and the contribution of the soft mode at $k_0$ to the observables is lost. This pattern is expected to repeat as $k_0$ varies with $\bar{h}$ resulting in the oscillations.  This is also illustrated in Fig.~\ref{fig:k0}, where we plot both $k_0$ and the $k_n$ for $N=12$. For larger system sizes, the probability that $k_0$ is near a $k_n$ is higher and the period and amplitude of the oscillations decrease.

To summarize, the oscillations of $g^{(2)}_{1N}$ essentially pick up the tendency of the underlying spin system to exhibit quasi long range order at $k_0$ as the transition boundary is approached both from above and below. In this way the finite size oscillations can be viewed as a direct probe of magnetic ordering in a finite sized system.\\

\subsection{Three-qubit toy model}

Here we derive an approximate analytic solution for the  NESS observables  discussed in the preceding section for the special case of
the Ising model in a transverse field (TFIM) with $J_{yi}=0$ in Eq.~(\ref{model}), corresponding to the horizontal line in Fig.~\ref{fig:pd}(a) with $\gamma=1$. 
We obtain a simple toy model for the finite size system, by projecting out all excited states except for the two lowest ones, which correspond to the finite-size split FM states in the FM phase (see Appendix~\ref{app: exact solution}). The interaction of the auxiliary qubits with this subspace is governed by the three-qubit Hamiltonian
\begin{eqnarray}
H_{\text{3-spin}} &=&\epsilon_\kappa \sigma^+_{\mathrm{M}}\sigma^-_{\mathrm{M}} + \bar{J}_{x,1}\sigma^x_1\sigma^x_{\mathrm{M}} + \bar{J}_{x,N-1}\sigma^x_N\sigma^x_{\mathrm{M}}\nonumber\,,
\end{eqnarray}
where the $\sigma_{\mathrm{M}}$-operators act on the two lowest states of the chain which have an energy splitting $\epsilon_\kappa$. The coupling coefficients between the auxiliary qubits and this low-energy subspace are renormalized to 
\begin{eqnarray}
\bar{J}_{x,1}=\phi_{\kappa,1} J_{x,1}\quad\mbox{and}\quad  \bar{J}_{x,N-1}=\psi_{\kappa,N-2} J_{x,N-1}\,,
\end{eqnarray}
where $\phi_{\kappa,1}$ and $\psi_{\kappa,N-2}$ correspond to the weights of the eigenstates of a TFIM with length $N-2$, which are given in Appendix~\ref{app: exact solution}, Eq.~(\ref{weights}).\\

The non-equilibrium dynamics of the 3-spin model are obtained from the master equation
\begin{equation}
\frac{\mathrm{d}\rho}{\mathrm{d}t}= -i \left[H_{\text{3-spin}}, \rho\right] +\sum_{\mu=1,N}\left(\Gamma^\downarrow_\mu \mathcal{D}[\sigma^-_\mu]\rho + \Gamma^\uparrow_\mu \mathcal{D}[\sigma^+_\mu]\rho\right)\,.
\end{equation}
We again use the third quantization formalism of Sec.~\ref{sec: third quantization} to solve this toy model. 
For the 3 spin model, the associated $\mathbf{X}$ and $\mathbf{Y}$ matrices defined in Eq.~(\ref{xymatrices}) for the correlation matrix are $6\times 6$ and largely sparse. Solving the Lyapunov equation~(\ref{lyapunov}) for the case of equal couplings $J_{x,1}=J_{x,N-1}$, equal dissipation rates $\Gamma\equiv\Gamma_{1}=\Gamma_{N}$ and $\Gamma_\pm\equiv\Gamma_{1}^{\pm}=\Gamma_{N}^{\pm}$
 and thermal photon numbers $n_{\rm th}\equiv n_1=n_N$, we obtain  
\begin{eqnarray}
\left\langle \sigma _{1,N}^{z}\right\rangle  &=&\frac{\Gamma _{-}\Gamma_{+}\left( {\bar{J}_{+}}^{2}+\epsilon_{\kappa}^{2}\right) }{{\bar{J}_{+}}^4+\Gamma _{+}^{2}\epsilon_{\kappa}^{2}}\,, \label{eq:sz} 
\end{eqnarray}
and 
\begin{equation}
\label{eq:g2}
g_{1N}^{(2)}=1+\frac{4\Gamma_-^2\bar{J}_x^4\epsilon_\kappa^2}{({\bar{J}_{+}}^{4}+\Gamma_-\Gamma_+{\bar{J}_{+}}^{2} + \Gamma_+(\Gamma_+ + \Gamma_-)\epsilon_\kappa^2)^2},
\end{equation}%
where $\bar{J}_{+}^2=2\bar{J}_{x}^2+\Gamma_+^2$ and $\bar{J}_x=\bar{J}_{x,1}=\bar{J}_{x,N-1}$.

Deep in the PM phase with $\bar{h}\rightarrow\infty$, the auxiliary qubits are decoupled from the qubit formed by the two lowest states of the chain, reflected by $\bar{J}_x\rightarrow 0$. 
 They are driven incoherently by the thermal population of the two transmission lines yielding
\begin{eqnarray}
\left\langle \sigma _{1,N}^{z}\right\rangle \approx - \frac{1}{2n_{\rm th} +1}\quad\mbox{for}\quad(\bar{h}\rightarrow\infty)\,.
\end{eqnarray}
In the high temperature limit, where $n_{\rm th}\rightarrow\infty$, $\langle\sigma _{1,N}^{z}\rangle=0$ and  $g_{1N}^{(2)}=1$. Low bath temperatures are therefore necessary to distinguish between different phases.
In the FM phase with $\bar{h}\rightarrow 0$, the auxiliary qubits are coupled to the qubit formed by the two lowest states of the chain with strength $\bar{J}_x\approx J_{x,1}$, but the energy splitting of the latter goes to zero, i.e., $\epsilon_\kappa \rightarrow 0$, such that
\begin{eqnarray}
\left\langle \sigma _{1,N}^{z}\right\rangle \approx \frac{\Gamma _{-}\Gamma
_{+}}{2 \bar{J}_{x}^{2}+\Gamma _{+}^{2}}\quad\mbox{for}\quad(\bar{h}\rightarrow 0)\,.
\end{eqnarray}
Consequently, the auxiliary qubits become fully saturated with $\left\langle \sigma _{1,N}^{z}\right\rangle \approx 0$ deep in the FM phase, provided $\bar{J}_x \gg \Gamma_+=\Gamma(2n_{\rm th}+1)$ in agreement with Fig.~\ref{fig:pd}(a).

The second-order correlation function in Eq.~(\ref{eq:g2}) is  trivial  in both phases since either $\epsilon_\kappa \rightarrow 0$ (for $\bar{h}\rightarrow 0$) 
or $\bar{J}_x\rightarrow 0$ (for $\bar{h}\rightarrow \infty$). However, non-trivial correlations arise for finite size systems at the phase boundaries where 
the product $\bar{J}_x \epsilon_\kappa$ is finite. Consequently $g_{1N}^{(2)}$ features a peak at $\bar{h}\approx1$ in agreement with the full numerical results in Fig.~\ref{fig:g2}. However, peak height and width as well as the center of the peak depend sensitively on the temperature and the size of the chain. In particular, for very low temperatures and dissipation rates, the peak can also be centered in the nominal FM phase depending on the precise choice of parameters. In general, larger temperature tends to broaden the peak and decreases the peak height, while larger system sizes tend to narrow the peak and increase the peak height.\\

\subsection{3SI Model}
\begin{figure}[t]
\begin{center}
\includegraphics[width=0.235\textwidth,clip]{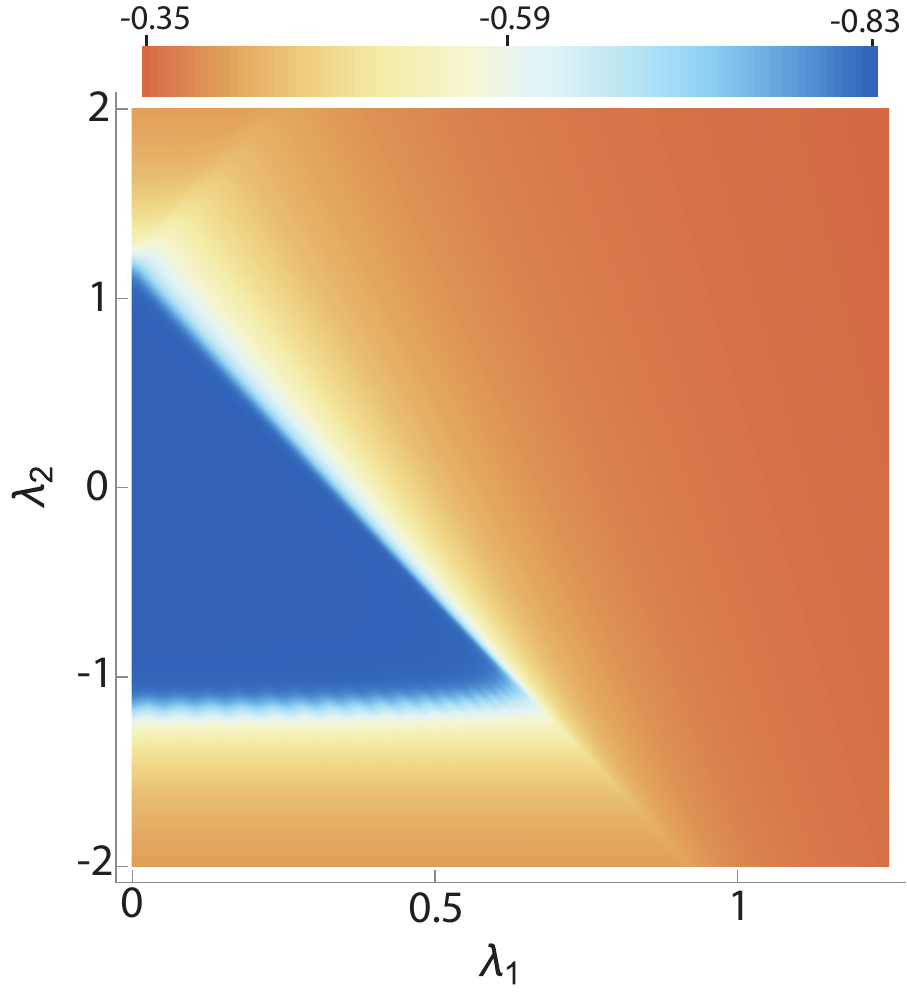}
\includegraphics[width=0.235\textwidth,clip]{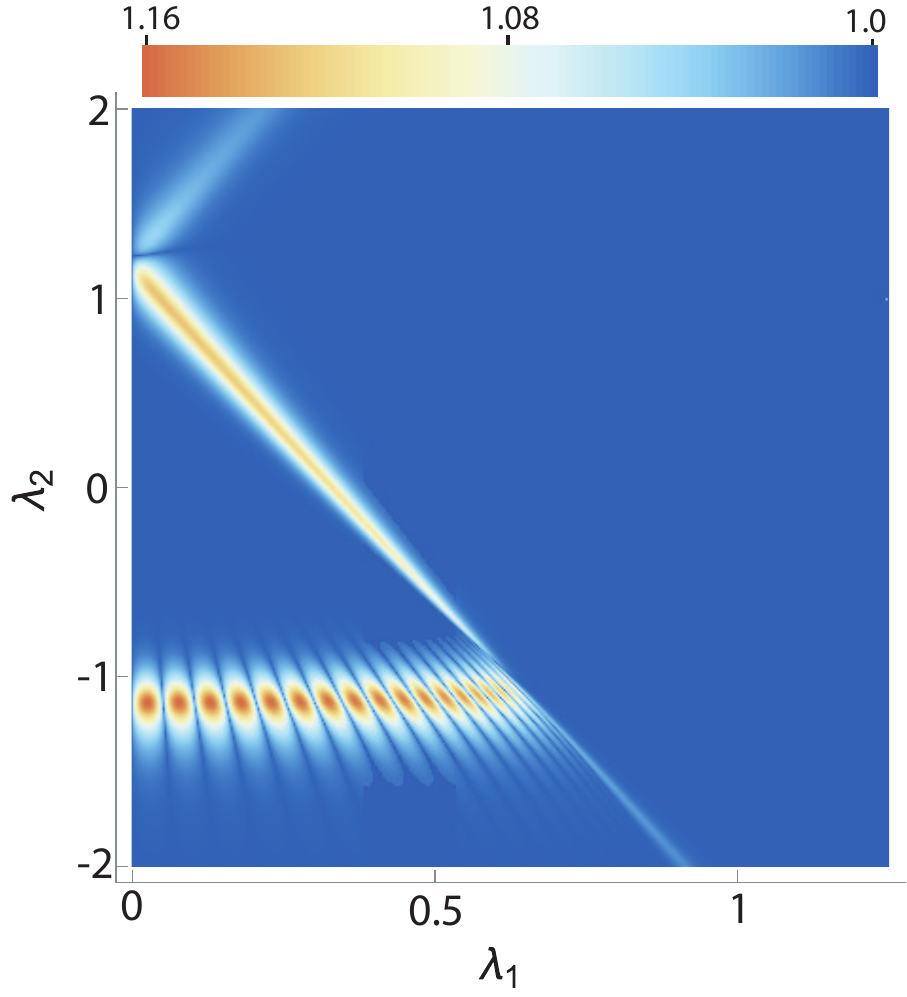}
\caption{
Auxiliary qubit occupation $\langle \sigma^z_1 \rangle$ (left panel) and second-order cross correlation function (right panel) in the NESS of the dual model (\ref{dual}) with dissipation from the boundaries. The system size is $N=42$.
}
\label{fig:dual}
\end{center}
\end{figure}
To explore the question of how sensitive the observables are to topological phase transitions, we study the related dual model of the  TXY model defined in Eq.~(\ref{dual}).
In the spin representation, this model has a topological (SPT), a non-topological (PM), and a symmetry breaking (FM) phase.
In the fermionic representation, as opposed to the TXY model, the dual model has three different phases with $n=0,1,2$ Majorana modes at each end of the open chain. 
The nature of the Majorana modes in different parts of the phase diagram varies and has been discussed in \cite{Niu2012}.  In zones with a single Majorana mode, the mode decays exponentially in the bulk without any oscillatory features.  However, in zones with two modes per boundary, one or both of the modes can be oscillatory in space. 

Given our results on the TXY model, it is imperative to ask the following two questions about the dual model:
i) Can the observables studied in this paper detect all the phase transitions -- topological as well as non-topological ones -- in the 3SI model? 
ii) Can these observables distinguish between topological and symmetry broken phases?
We find that the answer to question i) is positive, while that to question ii) is negative.

As shown in Fig.~\ref{fig:dual}, we find that the photon cross correlator for small finite size systems is indeed sensitive to transitions between all phases. The oscillatory behavior seen in the results for  $\lambda_2\sim-1$ and $\lambda_1 < 2$ for the transition between the $n=0$ and $n=2$ sectors stems from the interplay between finite size effects and the closure of a gap at incommensurate values of $k \neq 0 ,\pi$ as discussed in Sec.~\ref{sec: XY model}.

However, the situation is more complex regarding transitions between the $n=1$ and $n=2$ phases. By analyzing the Majorana mode amplitudes, we find that in zones with two modes per boundary, both observables couple to both Majorana end modes. However, one of the oscillatory modes has diminished amplitude at the site of the first spin and hence only couples weakly to the end cavity modes. This coupling gets weaker as the system size increases. Consequently, as one crosses a boundary from a region with $n=1$ to $n=2$, both observables show a weaker signature of this topological transition as compared to a transition from $n=0$ to $n=1$. 

\section{Conclusion}
\label{sec: conclusion}
In this paper, we  have explored  the possibility of probing ground state topological phenomena, especially Majorana bound states
in  a non equilibrium steady state of driven dissipative systems. We studied  two one dimensional spin models or equivalently,  models of spinless fermions known to exhibit nontrivial
topological features, subject to dissipation at the ends of the chain.    We find that though dissipation typically destroys topological protection,
local as well as bi-local observables studied in this paper  are able to detect  both symmetry breaking transitions as well as topological transitions between SPT phases exhibited by the non-dissipative model.   These observables, however, cannot distinguish between symmetry breaking and purely topological phase transitions.  
Combined with an a-priori knowledge of  the  ground states of the system, 
these observables can be perceived as directly probing changes in topology, i.e.,  detecting changes in the number of Majorana edge modes in a dissipative system.  We reiterate that, in the case of a fermionic chain, the bi-local correlator directly couples to the weights of the Majorana modes at the
 end of the chain.
 
 We have also discussed  a way to quantum engineer these  interesting spin models  with tunable parameters using superconducting
 qubits as well as extended the use of the auxiliary qubit spectroscopy discussed in Ref.~\cite{Bardyn2012}  to measure  the observables 
 calculated earlier.  
The strength of cavity-QED setups lies in the deliberate control that allows to construct specific models, and to directly measure correlation functions. This can be used to simulate Hamiltonians like that of the Kitaev wire or to build models for more general SPT phases, both of which we demonstrated in this work. Especially for the latter, which do not always feature topological end states, measuring observables that reveal the topological character could be very challenging in traditional condensed matter setups.

Our work opens up avenues for further research. It would be interesting to generalize the study here to two dimensional Kitaev models. In particular, it would be interesting to  define an observable that is sensitive to the topology of a state and  differentiate between phases far away from transition boundaries.  A good candidate is heat transport,
especially in two dimensional systems, where the Majorana edge modes offer an optimal path for transport, while the gapped bulk inhibits transport.  We leave these questions for future work.

\begin{appendix}

\section{Exact solution for the transverse-field Ising model with open boundaries}
\label{app: exact solution}
The TFIM is obtained from Eq.~(\ref{model}) with zero coupling in $y$-direction ($J_{y,i}=0$) and isotropic couplings in $x$-direction ($J_{x,i}=J_x$), i.e.,
\begin{eqnarray}
H_{\rm TFIM}&=& 2h \sum_i c_i^\dagger c_i  + J_x \sum_i (c_i^\dagger c_{i+1} +c_i^\dagger c_{i+1}^\dagger + \text{h.c.})\,.\nonumber\\
\end{eqnarray}
By introducing new fermionic operators
\begin{eqnarray}
\label{bogoliubov}
\eta_k=&\,
\frac12\sum_{j=1}^N
\left[
(\phi_{k,j}+\psi_{k,j})c_j+(\phi_{k,j}-\psi_{k,j})c^\dagger_j
\right]
\end{eqnarray}
or equivalently
\begin{eqnarray}
c_j=&\,
\frac12\sum_{k\geq0}
\left[
(\phi_{k,j}+\psi_{k,j})\eta_k+(\phi_{k,j}-\psi_{k,j})\eta^\dagger_k
\right]
\end{eqnarray}
one obtains a diagonalised Hamiltonian
\begin{eqnarray}
H_{\mathrm{TFIM}} = \sum_k \epsilon_k (\eta_k^\dagger \eta_k -1/2)
\end{eqnarray}
with energy spectrum
\begin{eqnarray}
\label{spectrum}
\epsilon_k  =2\sqrt{h^2+J_x^2 + 2hJ_x \cos k}\,.
\end{eqnarray}
In (\ref{bogoliubov}) the weights are chosen as
\begin{eqnarray}
\label{weights}
\phi_{k,j}&=& A_k \sin k(N+1-j)\\  
\psi_{k,j}&=&- \mathrm{sign} \Big[ \dfrac{J_x \sin k}{\sin k(N+1)} \Big] A_k \sin k j
\end{eqnarray}
with
\begin{eqnarray}
A_k  &= 2 \big(2N+1- \sin [k(2N+1)]/\sin k \big)^{-1/2}\,.
\end{eqnarray}
Here, $\xi = -h/J_x$ is the reduced transverse field, and the possible values of $k$ are solutions of 
\ma{
\dfrac{\sin k N}{\sin k (N+1)} = \xi. \label{kvalues}
}
For large transverse fields with $|\xi| \geq N/(N+1)$, Eq.\eqref{kvalues} has $N$ real solutions in the interval $ [0,\pi]$. 
However, for small fields, i.e., $|\xi| < N/(N+1)$, there is also one imaginary solution $k^\prime =\mathrm{i} \kappa$ with 
$k^\prime= \pi + \mathrm{i} \kappa$ for positive (negative) $\xi$ with $\sinh \kappa N/ \sinh \kappa (N+1) = |\xi|$.
It is well known that the imaginary solution $|1\rangle=\eta^\dagger_\kappa|0\rangle$ together with the ground state $|0\rangle$ correspond to a Majorana bound state localized at both edges of the chain. The latter forms the basis of the Majorana qubit with Pauli operators
\begin{eqnarray}
\label{qubit2}
\sigma^z_M = |1\rangle\langle 1| - |0\rangle\langle 0|\quad\quad \sigma^+_M=|1\rangle\langle 0|\,.
\end{eqnarray}
By projecting the full Hamiltonian on this low energy subspace we obtain the Majorana qubit Hamiltonian of the TFIM
\begin{eqnarray}
H_M=PH_{\rm TFIM} P = \epsilon_\kappa \sigma^z_M
\end{eqnarray}
with $P=|0\rangle\langle 0| + |1\rangle\langle 1|$ and vanishing energy splitting $\epsilon_{\kappa} \rightarrow 0$ if $|\xi|\rightarrow 0$ or $N\rightarrow \infty$.
\end{appendix}

\begin{acknowledgments}
We  thank, Atac Imamoglu, Charles-Edouard Bardyn and  Curt von Keyserlingk for enlightening discussions.
YD thanks the National Natural Science Foundation of China, under Grants No. 11204197, and No. 11474211.
\end{acknowledgments}

\bibliography{majorana_bib}

\begin{thebibliography}{37}%
\makeatletter
\providecommand \@ifxundefined [1]{%
 \@ifx{#1\undefined}
}%
\providecommand \@ifnum [1]{%
 \ifnum #1\expandafter \@firstoftwo
 \else \expandafter \@secondoftwo
 \fi
}%
\providecommand \@ifx [1]{%
 \ifx #1\expandafter \@firstoftwo
 \else \expandafter \@secondoftwo
 \fi
}%
\providecommand \natexlab [1]{#1}%
\providecommand \enquote  [1]{``#1''}%
\providecommand \bibnamefont  [1]{#1}%
\providecommand \bibfnamefont [1]{#1}%
\providecommand \citenamefont [1]{#1}%
\providecommand \href@noop [0]{\@secondoftwo}%
\providecommand \href [0]{\begingroup \@sanitize@url \@href}%
\providecommand \@href[1]{\@@startlink{#1}\@@href}%
\providecommand \@@href[1]{\endgroup#1\@@endlink}%
\providecommand \@sanitize@url [0]{\catcode `\\12\catcode `\$12\catcode
  `\&12\catcode `\#12\catcode `\^12\catcode `\_12\catcode `\%12\relax}%
\providecommand \@@startlink[1]{}%
\providecommand \@@endlink[0]{}%
\providecommand \url  [0]{\begingroup\@sanitize@url \@url }%
\providecommand \@url [1]{\endgroup\@href {#1}{\urlprefix }}%
\providecommand \urlprefix  [0]{URL }%
\providecommand \Eprint [0]{\href }%
\providecommand \doibase [0]{http://dx.doi.org/}%
\providecommand \selectlanguage [0]{\@gobble}%
\providecommand \bibinfo  [0]{\@secondoftwo}%
\providecommand \bibfield  [0]{\@secondoftwo}%
\providecommand \translation [1]{[#1]}%
\providecommand \BibitemOpen [0]{}%
\providecommand \bibitemStop [0]{}%
\providecommand \bibitemNoStop [0]{.\EOS\space}%
\providecommand \EOS [0]{\spacefactor3000\relax}%
\providecommand \BibitemShut  [1]{\csname bibitem#1\endcsname}%
\let\auto@bib@innerbib\@empty
\bibitem [{\citenamefont {Hasan}\ and\ \citenamefont {Kane}(2010)}]{Hasan2010}%
  \BibitemOpen
  \bibfield  {author} {\bibinfo {author} {\bibfnamefont {M.~Z.}\ \bibnamefont
  {Hasan}}\ and\ \bibinfo {author} {\bibfnamefont {C.~L.}\ \bibnamefont
  {Kane}},\ }\href {\doibase 10.1103/RevModPhys.82.3045} {\bibfield  {journal}
  {\bibinfo  {journal} {Rev. Mod. Phys.}\ }\textbf {\bibinfo {volume} {82}},\
  \bibinfo {pages} {3045} (\bibinfo {year} {2010})}\BibitemShut {NoStop}%
\bibitem [{\citenamefont {Qi}\ and\ \citenamefont {Zhang}(2011)}]{Qi2011}%
  \BibitemOpen
  \bibfield  {author} {\bibinfo {author} {\bibfnamefont {X.-L.}\ \bibnamefont
  {Qi}}\ and\ \bibinfo {author} {\bibfnamefont {S.-C.}\ \bibnamefont {Zhang}},\
  }\href {\doibase 10.1103/RevModPhys.83.1057} {\bibfield  {journal} {\bibinfo
  {journal} {Rev. Mod. Phys.}\ }\textbf {\bibinfo {volume} {83}},\ \bibinfo
  {pages} {1057} (\bibinfo {year} {2011})}\BibitemShut {NoStop}%
\bibitem [{\citenamefont {Kitaev}(2001)}]{Kitaev2001}%
  \BibitemOpen
  \bibfield  {author} {\bibinfo {author} {\bibfnamefont {A.~Y.}\ \bibnamefont
  {Kitaev}},\ }\href {http://stacks.iop.org/1063-7869/44/i=10S/a=S29}
  {\bibfield  {journal} {\bibinfo  {journal} {Physics-Uspekhi}\ }\textbf
  {\bibinfo {volume} {44}},\ \bibinfo {pages} {131} (\bibinfo {year}
  {2001})}\BibitemShut {NoStop}%
\bibitem [{\citenamefont {Nayak}\ \emph {et~al.}(2008)\citenamefont {Nayak},
  \citenamefont {Simon}, \citenamefont {Stern}, \citenamefont {Freedman},\ and\
  \citenamefont {Das~Sarma}}]{Nayak2008}%
  \BibitemOpen
  \bibfield  {author} {\bibinfo {author} {\bibfnamefont {C.}~\bibnamefont
  {Nayak}}, \bibinfo {author} {\bibfnamefont {S.~H.}\ \bibnamefont {Simon}},
  \bibinfo {author} {\bibfnamefont {A.}~\bibnamefont {Stern}}, \bibinfo
  {author} {\bibfnamefont {M.}~\bibnamefont {Freedman}}, \ and\ \bibinfo
  {author} {\bibfnamefont {S.}~\bibnamefont {Das~Sarma}},\ }\href {\doibase
  10.1103/RevModPhys.80.1083} {\bibfield  {journal} {\bibinfo  {journal} {Rev.
  Mod. Phys.}\ }\textbf {\bibinfo {volume} {80}},\ \bibinfo {pages} {1083}
  (\bibinfo {year} {2008})}\BibitemShut {NoStop}%
\bibitem [{\citenamefont {Lutchyn}\ \emph {et~al.}(2010)\citenamefont
  {Lutchyn}, \citenamefont {Sau},\ and\ \citenamefont
  {Das~Sarma}}]{Lutchyn2010}%
  \BibitemOpen
  \bibfield  {author} {\bibinfo {author} {\bibfnamefont {R.~M.}\ \bibnamefont
  {Lutchyn}}, \bibinfo {author} {\bibfnamefont {J.~D.}\ \bibnamefont {Sau}}, \
  and\ \bibinfo {author} {\bibfnamefont {S.}~\bibnamefont {Das~Sarma}},\ }\href
  {\doibase 10.1103/PhysRevLett.105.077001} {\bibfield  {journal} {\bibinfo
  {journal} {Phys. Rev. Lett.}\ }\textbf {\bibinfo {volume} {105}},\ \bibinfo
  {pages} {077001} (\bibinfo {year} {2010})}\BibitemShut {NoStop}%
\bibitem [{\citenamefont {Alicea}(2012)}]{Alicea2012}%
  \BibitemOpen
  \bibfield  {author} {\bibinfo {author} {\bibfnamefont {J.}~\bibnamefont
  {Alicea}},\ }\href {http://stacks.iop.org/0034-4885/75/i=7/a=076501}
  {\bibfield  {journal} {\bibinfo  {journal} {Reports on Progress in Physics}\
  }\textbf {\bibinfo {volume} {75}},\ \bibinfo {pages} {076501} (\bibinfo
  {year} {2012})}\BibitemShut {NoStop}%
\bibitem [{\citenamefont {Lu}\ \emph {et~al.}(2014)\citenamefont {Lu},
  \citenamefont {Joannopoulos},\ and\ \citenamefont {Soljacic}}]{Lu2014}%
  \BibitemOpen
  \bibfield  {author} {\bibinfo {author} {\bibfnamefont {L.}~\bibnamefont
  {Lu}}, \bibinfo {author} {\bibfnamefont {J.~D.}\ \bibnamefont
  {Joannopoulos}}, \ and\ \bibinfo {author} {\bibfnamefont {M.}~\bibnamefont
  {Soljacic}},\ }\href {http://dx.doi.org/10.1038/nphoton.2014.248} {\bibfield
  {journal} {\bibinfo  {journal} {Nat Photon}\ }\textbf {\bibinfo {volume}
  {8}},\ \bibinfo {pages} {821} (\bibinfo {year} {2014})}\BibitemShut {NoStop}%
\bibitem [{\citenamefont {Wang}\ \emph {et~al.}(2009)\citenamefont {Wang},
  \citenamefont {Chong}, \citenamefont {Joannopoulos},\ and\ \citenamefont
  {Soljacic}}]{Wang2009}%
  \BibitemOpen
  \bibfield  {author} {\bibinfo {author} {\bibfnamefont {Z.}~\bibnamefont
  {Wang}}, \bibinfo {author} {\bibfnamefont {Y.}~\bibnamefont {Chong}},
  \bibinfo {author} {\bibfnamefont {J.~D.}\ \bibnamefont {Joannopoulos}}, \
  and\ \bibinfo {author} {\bibfnamefont {M.}~\bibnamefont {Soljacic}},\ }\href
  {http://dx.doi.org/10.1038/nature08293} {\bibfield  {journal} {\bibinfo
  {journal} {Nature}\ }\textbf {\bibinfo {volume} {461}},\ \bibinfo {pages}
  {772} (\bibinfo {year} {2009})}\BibitemShut {NoStop}%
\bibitem [{\citenamefont {M.}\ \emph {et~al.}(2013)\citenamefont {M.},
  \citenamefont {S.}, \citenamefont {J.}, \citenamefont {A.},\ and\
  \citenamefont {M.}}]{Hafezi2013}%
  \BibitemOpen
  \bibfield  {author} {\bibinfo {author} {\bibfnamefont {H.}~\bibnamefont
  {M.}}, \bibinfo {author} {\bibfnamefont {M.}~\bibnamefont {S.}}, \bibinfo
  {author} {\bibfnamefont {F.}~\bibnamefont {J.}}, \bibinfo {author}
  {\bibfnamefont {M.}~\bibnamefont {A.}}, \ and\ \bibinfo {author}
  {\bibfnamefont {T.~J.}\ \bibnamefont {M.}},\ }\href
  {http://dx.doi.org/10.1038/nphoton.2013.274} {\bibfield  {journal} {\bibinfo
  {journal} {Nat Photon}\ }\textbf {\bibinfo {volume} {7}},\ \bibinfo {pages}
  {1001} (\bibinfo {year} {2013})}\BibitemShut {NoStop}%
\bibitem [{\citenamefont {Rechtsman}\ \emph {et~al.}(2013)\citenamefont
  {Rechtsman}, \citenamefont {Zeuner}, \citenamefont {Plotnik}, \citenamefont
  {Lumer}, \citenamefont {Podolsky}, \citenamefont {Dreisow}, \citenamefont
  {Nolte}, \citenamefont {Segev},\ and\ \citenamefont
  {Szameit}}]{Rechtsman2013}%
  \BibitemOpen
  \bibfield  {author} {\bibinfo {author} {\bibfnamefont {M.~C.}\ \bibnamefont
  {Rechtsman}}, \bibinfo {author} {\bibfnamefont {J.~M.}\ \bibnamefont
  {Zeuner}}, \bibinfo {author} {\bibfnamefont {Y.}~\bibnamefont {Plotnik}},
  \bibinfo {author} {\bibfnamefont {Y.}~\bibnamefont {Lumer}}, \bibinfo
  {author} {\bibfnamefont {D.}~\bibnamefont {Podolsky}}, \bibinfo {author}
  {\bibfnamefont {F.}~\bibnamefont {Dreisow}}, \bibinfo {author} {\bibfnamefont
  {S.}~\bibnamefont {Nolte}}, \bibinfo {author} {\bibfnamefont
  {M.}~\bibnamefont {Segev}}, \ and\ \bibinfo {author} {\bibfnamefont
  {A.}~\bibnamefont {Szameit}},\ }\href {http://dx.doi.org/10.1038/nature12066}
  {\bibfield  {journal} {\bibinfo  {journal} {Nature}\ }\textbf {\bibinfo
  {volume} {496}},\ \bibinfo {pages} {196} (\bibinfo {year}
  {2013})}\BibitemShut {NoStop}%
\bibitem [{\citenamefont {Vardeny}\ \emph {et~al.}(2013)\citenamefont
  {Vardeny}, \citenamefont {Nahata},\ and\ \citenamefont
  {Agrawal}}]{Vardeny2013}%
  \BibitemOpen
  \bibfield  {author} {\bibinfo {author} {\bibfnamefont {Z.~V.}\ \bibnamefont
  {Vardeny}}, \bibinfo {author} {\bibfnamefont {A.}~\bibnamefont {Nahata}}, \
  and\ \bibinfo {author} {\bibfnamefont {A.}~\bibnamefont {Agrawal}},\ }\href
  {http://dx.doi.org/10.1038/nphoton.2012.343} {\bibfield  {journal} {\bibinfo
  {journal} {Nat Photon}\ }\textbf {\bibinfo {volume} {7}},\ \bibinfo {pages}
  {177} (\bibinfo {year} {2013})}\BibitemShut {NoStop}%
\bibitem [{\citenamefont {Tanese}\ \emph {et~al.}(2014)\citenamefont {Tanese},
  \citenamefont {Gurevich}, \citenamefont {Baboux}, \citenamefont {Jacqmin},
  \citenamefont {Lema\^{\i}tre}, \citenamefont {Galopin}, \citenamefont
  {Sagnes}, \citenamefont {Amo}, \citenamefont {Bloch},\ and\ \citenamefont
  {Akkermans}}]{Tanese2014}%
  \BibitemOpen
  \bibfield  {author} {\bibinfo {author} {\bibfnamefont {D.}~\bibnamefont
  {Tanese}}, \bibinfo {author} {\bibfnamefont {E.}~\bibnamefont {Gurevich}},
  \bibinfo {author} {\bibfnamefont {F.}~\bibnamefont {Baboux}}, \bibinfo
  {author} {\bibfnamefont {T.}~\bibnamefont {Jacqmin}}, \bibinfo {author}
  {\bibfnamefont {A.}~\bibnamefont {Lema\^{\i}tre}}, \bibinfo {author}
  {\bibfnamefont {E.}~\bibnamefont {Galopin}}, \bibinfo {author} {\bibfnamefont
  {I.}~\bibnamefont {Sagnes}}, \bibinfo {author} {\bibfnamefont
  {A.}~\bibnamefont {Amo}}, \bibinfo {author} {\bibfnamefont {J.}~\bibnamefont
  {Bloch}}, \ and\ \bibinfo {author} {\bibfnamefont {E.}~\bibnamefont
  {Akkermans}},\ }\href {\doibase 10.1103/PhysRevLett.112.146404} {\bibfield
  {journal} {\bibinfo  {journal} {Phys. Rev. Lett.}\ }\textbf {\bibinfo
  {volume} {112}},\ \bibinfo {pages} {146404} (\bibinfo {year}
  {2014})}\BibitemShut {NoStop}%
\bibitem [{\citenamefont {Bardyn}\ and\ \citenamefont
  {Imamoglu}(2012)}]{Bardyn2012}%
  \BibitemOpen
  \bibfield  {author} {\bibinfo {author} {\bibfnamefont {C.-E.}\ \bibnamefont
  {Bardyn}}\ and\ \bibinfo {author} {\bibfnamefont {A.}~\bibnamefont
  {Imamoglu}},\ }\href {\doibase 10.1103/PhysRevLett.109.253606} {\bibfield
  {journal} {\bibinfo  {journal} {Phys. Rev. Lett.}\ }\textbf {\bibinfo
  {volume} {109}},\ \bibinfo {pages} {253606} (\bibinfo {year}
  {2012})}\BibitemShut {NoStop}%
\bibitem [{\citenamefont {Levitov}\ \emph {et~al.}(2001)\citenamefont
  {Levitov}, \citenamefont {Orlando}, \citenamefont {Majer},\ and\
  \citenamefont {Mooij}}]{Levitov2001}%
  \BibitemOpen
  \bibfield  {author} {\bibinfo {author} {\bibfnamefont {L.}~\bibnamefont
  {Levitov}}, \bibinfo {author} {\bibfnamefont {T.}~\bibnamefont {Orlando}},
  \bibinfo {author} {\bibfnamefont {J.}~\bibnamefont {Majer}}, \ and\ \bibinfo
  {author} {\bibfnamefont {J.}~\bibnamefont {Mooij}},\ }\href@noop {}
  {\bibfield  {journal} {\bibinfo  {journal} {arXiv preprint cond-mat/0108266}\
  } (\bibinfo {year} {2001})}\BibitemShut {NoStop}%
\bibitem [{\citenamefont {Zvyagin}(2013)}]{Zvyagin2013}%
  \BibitemOpen
  \bibfield  {author} {\bibinfo {author} {\bibfnamefont {A.~A.}\ \bibnamefont
  {Zvyagin}},\ }\href {\doibase 10.1103/PhysRevLett.110.217207} {\bibfield
  {journal} {\bibinfo  {journal} {Phys. Rev. Lett.}\ }\textbf {\bibinfo
  {volume} {110}},\ \bibinfo {pages} {217207} (\bibinfo {year}
  {2013})}\BibitemShut {NoStop}%
\bibitem [{\citenamefont {Hwang}\ and\ \citenamefont {Choi}(2013)}]{Hwang2013}%
  \BibitemOpen
  \bibfield  {author} {\bibinfo {author} {\bibfnamefont {M.-J.}\ \bibnamefont
  {Hwang}}\ and\ \bibinfo {author} {\bibfnamefont {M.-S.}\ \bibnamefont
  {Choi}},\ }\href {\doibase 10.1103/PhysRevB.87.125404} {\bibfield  {journal}
  {\bibinfo  {journal} {Phys. Rev. B}\ }\textbf {\bibinfo {volume} {87}},\
  \bibinfo {pages} {125404} (\bibinfo {year} {2013})}\BibitemShut {NoStop}%
\bibitem [{\citenamefont {Zhu}\ \emph {et~al.}(2013)\citenamefont {Zhu},
  \citenamefont {Schmidt},\ and\ \citenamefont {Koch}}]{Zhu2013}%
  \BibitemOpen
  \bibfield  {author} {\bibinfo {author} {\bibfnamefont {G.}~\bibnamefont
  {Zhu}}, \bibinfo {author} {\bibfnamefont {S.}~\bibnamefont {Schmidt}}, \ and\
  \bibinfo {author} {\bibfnamefont {J.}~\bibnamefont {Koch}},\ }\href
  {http://stacks.iop.org/1367-2630/15/i=11/a=115002} {\bibfield  {journal}
  {\bibinfo  {journal} {New Journal of Physics}\ }\textbf {\bibinfo {volume}
  {15}},\ \bibinfo {pages} {115002} (\bibinfo {year} {2013})}\BibitemShut
  {NoStop}%
\bibitem [{\citenamefont {Niu}\ \emph {et~al.}(2012)\citenamefont {Niu},
  \citenamefont {Chung}, \citenamefont {Hsu}, \citenamefont {Mandal},
  \citenamefont {Raghu},\ and\ \citenamefont {Chakravarty}}]{Niu2012}%
  \BibitemOpen
  \bibfield  {author} {\bibinfo {author} {\bibfnamefont {Y.}~\bibnamefont
  {Niu}}, \bibinfo {author} {\bibfnamefont {S.~B.}\ \bibnamefont {Chung}},
  \bibinfo {author} {\bibfnamefont {C.-H.}\ \bibnamefont {Hsu}}, \bibinfo
  {author} {\bibfnamefont {I.}~\bibnamefont {Mandal}}, \bibinfo {author}
  {\bibfnamefont {S.}~\bibnamefont {Raghu}}, \ and\ \bibinfo {author}
  {\bibfnamefont {S.}~\bibnamefont {Chakravarty}},\ }\href {\doibase
  10.1103/PhysRevB.85.035110} {\bibfield  {journal} {\bibinfo  {journal} {Phys.
  Rev. B}\ }\textbf {\bibinfo {volume} {85}},\ \bibinfo {pages} {035110}
  (\bibinfo {year} {2012})}\BibitemShut {NoStop}%
\bibitem [{\citenamefont {Eichler}\ \emph {et~al.}(2014)\citenamefont
  {Eichler}, \citenamefont {Salathe}, \citenamefont {Mlynek}, \citenamefont
  {Schmidt},\ and\ \citenamefont {Wallraff}}]{Eichler2014}%
  \BibitemOpen
  \bibfield  {author} {\bibinfo {author} {\bibfnamefont {C.}~\bibnamefont
  {Eichler}}, \bibinfo {author} {\bibfnamefont {Y.}~\bibnamefont {Salathe}},
  \bibinfo {author} {\bibfnamefont {J.}~\bibnamefont {Mlynek}}, \bibinfo
  {author} {\bibfnamefont {S.}~\bibnamefont {Schmidt}}, \ and\ \bibinfo
  {author} {\bibfnamefont {A.}~\bibnamefont {Wallraff}},\ }\href {\doibase
  10.1103/PhysRevLett.113.110502} {\bibfield  {journal} {\bibinfo  {journal}
  {Phys. Rev. Lett.}\ }\textbf {\bibinfo {volume} {113}},\ \bibinfo {pages}
  {110502} (\bibinfo {year} {2014})}\BibitemShut {NoStop}%
\bibitem [{\citenamefont {Prosen}(2008)}]{Prosen2008}%
  \BibitemOpen
  \bibfield  {author} {\bibinfo {author} {\bibfnamefont {T.}~\bibnamefont
  {Prosen}},\ }\href {http://stacks.iop.org/1367-2630/10/i=4/a=043026}
  {\bibfield  {journal} {\bibinfo  {journal} {New Journal of Physics}\ }\textbf
  {\bibinfo {volume} {10}},\ \bibinfo {pages} {043026} (\bibinfo {year}
  {2008})}\BibitemShut {NoStop}%
\bibitem [{\citenamefont {Dutta}\ \emph {et~al.}(2015)\citenamefont {Dutta},
  \citenamefont {Aeppli}, \citenamefont {Chakrabarti}, \citenamefont
  {Divakaran}, \citenamefont {Rosenbaum},\ and\ \citenamefont
  {Sen}}]{Dutta2015}%
  \BibitemOpen
  \bibfield  {author} {\bibinfo {author} {\bibfnamefont {A.}~\bibnamefont
  {Dutta}}, \bibinfo {author} {\bibfnamefont {G.}~\bibnamefont {Aeppli}},
  \bibinfo {author} {\bibfnamefont {B.~K.}\ \bibnamefont {Chakrabarti}},
  \bibinfo {author} {\bibfnamefont {U.}~\bibnamefont {Divakaran}}, \bibinfo
  {author} {\bibfnamefont {T.~F.}\ \bibnamefont {Rosenbaum}}, \ and\ \bibinfo
  {author} {\bibfnamefont {D.}~\bibnamefont {Sen}},\ }\href
  {https://cds.cern.ch/record/2017717} {\emph {\bibinfo {title} {{Quantum phase
  transitions in transverse field spin models: from statistical physics to
  quantum information}}}}\ (\bibinfo  {publisher} {Cambridge University
  Press},\ \bibinfo {address} {Cambridge},\ \bibinfo {year} {2015})\BibitemShut
  {NoStop}%
\bibitem [{\citenamefont {Schnyder}\ \emph {et~al.}(2008)\citenamefont
  {Schnyder}, \citenamefont {Ryu}, \citenamefont {Furusaki},\ and\
  \citenamefont {Ludwig}}]{Schnyder08}%
  \BibitemOpen
  \bibfield  {author} {\bibinfo {author} {\bibfnamefont {A.~P.}\ \bibnamefont
  {Schnyder}}, \bibinfo {author} {\bibfnamefont {S.}~\bibnamefont {Ryu}},
  \bibinfo {author} {\bibfnamefont {A.}~\bibnamefont {Furusaki}}, \ and\
  \bibinfo {author} {\bibfnamefont {A.~W.~W.}\ \bibnamefont {Ludwig}},\ }\href
  {\doibase 10.1103/PhysRevB.78.195125} {\bibfield  {journal} {\bibinfo
  {journal} {Phys. Rev. B}\ }\textbf {\bibinfo {volume} {78}},\ \bibinfo
  {pages} {195125} (\bibinfo {year} {2008})}\BibitemShut {NoStop}%
\bibitem [{\citenamefont {Ryu}\ \emph {et~al.}(2010)\citenamefont {Ryu},
  \citenamefont {Schnyder}, \citenamefont {Furusaki},\ and\ \citenamefont
  {Ludwig}}]{Ryu10}%
  \BibitemOpen
  \bibfield  {author} {\bibinfo {author} {\bibfnamefont {S.}~\bibnamefont
  {Ryu}}, \bibinfo {author} {\bibfnamefont {A.~P.}\ \bibnamefont {Schnyder}},
  \bibinfo {author} {\bibfnamefont {A.}~\bibnamefont {Furusaki}}, \ and\
  \bibinfo {author} {\bibfnamefont {A.~W.~W.}\ \bibnamefont {Ludwig}},\ }\href
  {http://stacks.iop.org/1367-2630/12/i=6/a=065010} {\bibfield  {journal}
  {\bibinfo  {journal} {New Journal of Physics}\ }\textbf {\bibinfo {volume}
  {12}},\ \bibinfo {pages} {065010} (\bibinfo {year} {2010})}\BibitemShut
  {NoStop}%
\bibitem [{\citenamefont {Chen}\ \emph {et~al.}(2013)\citenamefont {Chen},
  \citenamefont {Gu}, \citenamefont {Liu},\ and\ \citenamefont {Wen}}]{Chen13}%
  \BibitemOpen
  \bibfield  {author} {\bibinfo {author} {\bibfnamefont {X.}~\bibnamefont
  {Chen}}, \bibinfo {author} {\bibfnamefont {Z.-C.}\ \bibnamefont {Gu}},
  \bibinfo {author} {\bibfnamefont {Z.-X.}\ \bibnamefont {Liu}}, \ and\
  \bibinfo {author} {\bibfnamefont {X.-G.}\ \bibnamefont {Wen}},\ }\href
  {\doibase 10.1103/PhysRevB.87.155114} {\bibfield  {journal} {\bibinfo
  {journal} {Phys. Rev. B}\ }\textbf {\bibinfo {volume} {87}},\ \bibinfo
  {pages} {155114} (\bibinfo {year} {2013})}\BibitemShut {NoStop}%
\bibitem [{\citenamefont {Yao}\ \emph {et~al.}(2015)\citenamefont {Yao},
  \citenamefont {Laumann},\ and\ \citenamefont {Vishwanath}}]{Yao2015}%
  \BibitemOpen
  \bibfield  {author} {\bibinfo {author} {\bibfnamefont {N.~Y.}\ \bibnamefont
  {Yao}}, \bibinfo {author} {\bibfnamefont {C.~R.}\ \bibnamefont {Laumann}}, \
  and\ \bibinfo {author} {\bibfnamefont {A.}~\bibnamefont {Vishwanath}},\
  }\href {http://arxiv.org/abs/1508.06995} {\  (\bibinfo {year} {2015})},\
  \Eprint {http://arxiv.org/abs/arxiv:1508.06995} {arxiv:1508.06995}
  \BibitemShut {NoStop}%
\bibitem [{\citenamefont {Salath\'e}\ \emph {et~al.}(2015)\citenamefont
  {Salath\'e}, \citenamefont {Mondal}, \citenamefont {Oppliger}, \citenamefont
  {Heinsoo}, \citenamefont {Kurpiers}, \citenamefont {Potocnik}, \citenamefont
  {Mezzacapo}, \citenamefont {Las~Heras}, \citenamefont {Lamata}, \citenamefont
  {Solano}, \citenamefont {Filipp},\ and\ \citenamefont
  {Wallraff}}]{Salathe2015}%
  \BibitemOpen
  \bibfield  {author} {\bibinfo {author} {\bibfnamefont {Y.}~\bibnamefont
  {Salath\'e}}, \bibinfo {author} {\bibfnamefont {M.}~\bibnamefont {Mondal}},
  \bibinfo {author} {\bibfnamefont {M.}~\bibnamefont {Oppliger}}, \bibinfo
  {author} {\bibfnamefont {J.}~\bibnamefont {Heinsoo}}, \bibinfo {author}
  {\bibfnamefont {P.}~\bibnamefont {Kurpiers}}, \bibinfo {author}
  {\bibfnamefont {A.}~\bibnamefont {Potocnik}}, \bibinfo {author}
  {\bibfnamefont {A.}~\bibnamefont {Mezzacapo}}, \bibinfo {author}
  {\bibfnamefont {U.}~\bibnamefont {Las~Heras}}, \bibinfo {author}
  {\bibfnamefont {L.}~\bibnamefont {Lamata}}, \bibinfo {author} {\bibfnamefont
  {E.}~\bibnamefont {Solano}}, \bibinfo {author} {\bibfnamefont
  {S.}~\bibnamefont {Filipp}}, \ and\ \bibinfo {author} {\bibfnamefont
  {A.}~\bibnamefont {Wallraff}},\ }\href {\doibase 10.1103/PhysRevX.5.021027}
  {\bibfield  {journal} {\bibinfo  {journal} {Phys. Rev. X}\ }\textbf {\bibinfo
  {volume} {5}},\ \bibinfo {pages} {021027} (\bibinfo {year}
  {2015})}\BibitemShut {NoStop}%
\bibitem [{\citenamefont {Houck}\ \emph {et~al.}(2012)\citenamefont {Houck},
  \citenamefont {T\"{u}reci},\ and\ \citenamefont {Koch}}]{Houck2012}%
  \BibitemOpen
  \bibfield  {author} {\bibinfo {author} {\bibfnamefont {A.~A.}\ \bibnamefont
  {Houck}}, \bibinfo {author} {\bibfnamefont {H.~E.}\ \bibnamefont
  {T\"{u}reci}}, \ and\ \bibinfo {author} {\bibfnamefont {J.}~\bibnamefont
  {Koch}},\ }\href {\doibase 10.1038/nphys2251} {\bibfield  {journal} {\bibinfo
   {journal} {Nature Phys.}\ }\textbf {\bibinfo {volume} {8}},\ \bibinfo
  {pages} {292} (\bibinfo {year} {2012})}\BibitemShut {NoStop}%
\bibitem [{\citenamefont {Schmidt}\ and\ \citenamefont
  {Koch}(2013)}]{Schmidt2013*2}%
  \BibitemOpen
  \bibfield  {author} {\bibinfo {author} {\bibfnamefont {S.}~\bibnamefont
  {Schmidt}}\ and\ \bibinfo {author} {\bibfnamefont {J.}~\bibnamefont {Koch}},\
  }\href {\doibase 10.1002/andp.201200261} {\bibfield  {journal} {\bibinfo
  {journal} {Annalen der Physik}\ }\textbf {\bibinfo {volume} {525}},\ \bibinfo
  {pages} {395} (\bibinfo {year} {2013})}\BibitemShut {NoStop}%
\bibitem [{\citenamefont {Barends}\ \emph {et~al.}(2015)\citenamefont
  {Barends}, \citenamefont {Lamata}, \citenamefont {Kelly}, \citenamefont
  {Garcia-Alvarez}, \citenamefont {Fowler}, \citenamefont {Megrant},
  \citenamefont {Jeffrey}, \citenamefont {White}, \citenamefont {Sank},
  \citenamefont {Mutus}, \citenamefont {Campbell}, \citenamefont {Chen},
  \citenamefont {Chen}, \citenamefont {Chiaro}, \citenamefont {Dunsworth},
  \citenamefont {Hoi}, \citenamefont {Neill}, \citenamefont {O/'Malley},
  \citenamefont {Quintana}, \citenamefont {Roushan}, \citenamefont
  {Vainsencher}, \citenamefont {Wenner}, \citenamefont {Solano},\ and\
  \citenamefont {Martinis}}]{Barends2015}%
  \BibitemOpen
  \bibfield  {author} {\bibinfo {author} {\bibfnamefont {R.}~\bibnamefont
  {Barends}}, \bibinfo {author} {\bibfnamefont {L.}~\bibnamefont {Lamata}},
  \bibinfo {author} {\bibfnamefont {J.}~\bibnamefont {Kelly}}, \bibinfo
  {author} {\bibfnamefont {L.}~\bibnamefont {Garcia-Alvarez}}, \bibinfo
  {author} {\bibfnamefont {A.~G.}\ \bibnamefont {Fowler}}, \bibinfo {author}
  {\bibfnamefont {A.}~\bibnamefont {Megrant}}, \bibinfo {author} {\bibfnamefont
  {E.}~\bibnamefont {Jeffrey}}, \bibinfo {author} {\bibfnamefont {T.~C.}\
  \bibnamefont {White}}, \bibinfo {author} {\bibfnamefont {D.}~\bibnamefont
  {Sank}}, \bibinfo {author} {\bibfnamefont {J.~Y.}\ \bibnamefont {Mutus}},
  \bibinfo {author} {\bibfnamefont {B.}~\bibnamefont {Campbell}}, \bibinfo
  {author} {\bibfnamefont {Y.}~\bibnamefont {Chen}}, \bibinfo {author}
  {\bibfnamefont {Z.}~\bibnamefont {Chen}}, \bibinfo {author} {\bibfnamefont
  {B.}~\bibnamefont {Chiaro}}, \bibinfo {author} {\bibfnamefont
  {A.}~\bibnamefont {Dunsworth}}, \bibinfo {author} {\bibfnamefont {I.~C.}\
  \bibnamefont {Hoi}}, \bibinfo {author} {\bibfnamefont {C.}~\bibnamefont
  {Neill}}, \bibinfo {author} {\bibfnamefont {P.~J.~J.}\ \bibnamefont
  {O/'Malley}}, \bibinfo {author} {\bibfnamefont {C.}~\bibnamefont {Quintana}},
  \bibinfo {author} {\bibfnamefont {P.}~\bibnamefont {Roushan}}, \bibinfo
  {author} {\bibfnamefont {A.}~\bibnamefont {Vainsencher}}, \bibinfo {author}
  {\bibfnamefont {J.}~\bibnamefont {Wenner}}, \bibinfo {author} {\bibfnamefont
  {E.}~\bibnamefont {Solano}}, \ and\ \bibinfo {author} {\bibfnamefont {J.~M.}\
  \bibnamefont {Martinis}},\ }\href {http://dx.doi.org/10.1038/ncomms8654}
  {\bibfield  {journal} {\bibinfo  {journal} {Nat Commun}\ }\textbf {\bibinfo
  {volume} {6}} (\bibinfo {year} {2015})}\BibitemShut {NoStop}%
\bibitem [{\citenamefont {Chen}\ \emph {et~al.}(2014)\citenamefont {Chen},
  \citenamefont {Neill}, \citenamefont {Roushan}, \citenamefont {Leung},
  \citenamefont {Fang}, \citenamefont {Barends}, \citenamefont {Kelly},
  \citenamefont {Campbell}, \citenamefont {Chen}, \citenamefont {Chiaro},
  \citenamefont {Dunsworth}, \citenamefont {Jeffrey}, \citenamefont {Megrant},
  \citenamefont {Mutus}, \citenamefont {O'Malley}, \citenamefont {Quintana},
  \citenamefont {Sank}, \citenamefont {Vainsencher}, \citenamefont {Wenner},
  \citenamefont {White}, \citenamefont {Geller}, \citenamefont {Cleland},\ and\
  \citenamefont {Martinis}}]{Chen2014}%
  \BibitemOpen
  \bibfield  {author} {\bibinfo {author} {\bibfnamefont {Y.}~\bibnamefont
  {Chen}}, \bibinfo {author} {\bibfnamefont {C.}~\bibnamefont {Neill}},
  \bibinfo {author} {\bibfnamefont {P.}~\bibnamefont {Roushan}}, \bibinfo
  {author} {\bibfnamefont {N.}~\bibnamefont {Leung}}, \bibinfo {author}
  {\bibfnamefont {M.}~\bibnamefont {Fang}}, \bibinfo {author} {\bibfnamefont
  {R.}~\bibnamefont {Barends}}, \bibinfo {author} {\bibfnamefont
  {J.}~\bibnamefont {Kelly}}, \bibinfo {author} {\bibfnamefont
  {B.}~\bibnamefont {Campbell}}, \bibinfo {author} {\bibfnamefont
  {Z.}~\bibnamefont {Chen}}, \bibinfo {author} {\bibfnamefont {B.}~\bibnamefont
  {Chiaro}}, \bibinfo {author} {\bibfnamefont {A.}~\bibnamefont {Dunsworth}},
  \bibinfo {author} {\bibfnamefont {E.}~\bibnamefont {Jeffrey}}, \bibinfo
  {author} {\bibfnamefont {A.}~\bibnamefont {Megrant}}, \bibinfo {author}
  {\bibfnamefont {J.~Y.}\ \bibnamefont {Mutus}}, \bibinfo {author}
  {\bibfnamefont {P.~J.~J.}\ \bibnamefont {O'Malley}}, \bibinfo {author}
  {\bibfnamefont {C.~M.}\ \bibnamefont {Quintana}}, \bibinfo {author}
  {\bibfnamefont {D.}~\bibnamefont {Sank}}, \bibinfo {author} {\bibfnamefont
  {A.}~\bibnamefont {Vainsencher}}, \bibinfo {author} {\bibfnamefont
  {J.}~\bibnamefont {Wenner}}, \bibinfo {author} {\bibfnamefont {T.~C.}\
  \bibnamefont {White}}, \bibinfo {author} {\bibfnamefont {M.~R.}\ \bibnamefont
  {Geller}}, \bibinfo {author} {\bibfnamefont {A.~N.}\ \bibnamefont {Cleland}},
  \ and\ \bibinfo {author} {\bibfnamefont {J.~M.}\ \bibnamefont {Martinis}},\
  }\href {\doibase 10.1103/PhysRevLett.113.220502} {\bibfield  {journal}
  {\bibinfo  {journal} {Phys. Rev. Lett.}\ }\textbf {\bibinfo {volume} {113}},\
  \bibinfo {pages} {220502} (\bibinfo {year} {2014})}\BibitemShut {NoStop}%
\bibitem [{\citenamefont {Devoret}(1995)}]{Devoret1995}%
  \BibitemOpen
  \bibfield  {author} {\bibinfo {author} {\bibfnamefont {M.~H.}\ \bibnamefont
  {Devoret}},\ }\href
  {http://www.physique.usherb.ca/tremblay/cours/PHY-731/Quantum\_circuit\_theory-1.pdf}
  {\bibfield  {journal} {\bibinfo  {journal} {Les Houches, Session LXIII}\ }
  (\bibinfo {year} {1995})}\BibitemShut {NoStop}%
\bibitem [{\citenamefont {Koch}\ \emph {et~al.}(2007)\citenamefont {Koch},
  \citenamefont {Yu}, \citenamefont {Gambetta}, \citenamefont {Houck},
  \citenamefont {Schuster}, \citenamefont {Majer}, \citenamefont {Blais},
  \citenamefont {Devoret}, \citenamefont {Girvin},\ and\ \citenamefont
  {Schoelkopf}}]{Koch2007}%
  \BibitemOpen
  \bibfield  {author} {\bibinfo {author} {\bibfnamefont {J.}~\bibnamefont
  {Koch}}, \bibinfo {author} {\bibfnamefont {T.}~\bibnamefont {Yu}}, \bibinfo
  {author} {\bibfnamefont {J.}~\bibnamefont {Gambetta}}, \bibinfo {author}
  {\bibfnamefont {A.}~\bibnamefont {Houck}}, \bibinfo {author} {\bibfnamefont
  {D.}~\bibnamefont {Schuster}}, \bibinfo {author} {\bibfnamefont
  {J.}~\bibnamefont {Majer}}, \bibinfo {author} {\bibfnamefont
  {A.}~\bibnamefont {Blais}}, \bibinfo {author} {\bibfnamefont
  {M.}~\bibnamefont {Devoret}}, \bibinfo {author} {\bibfnamefont
  {S.}~\bibnamefont {Girvin}}, \ and\ \bibinfo {author} {\bibfnamefont
  {R.}~\bibnamefont {Schoelkopf}},\ }\href {\doibase
  10.1103/PhysRevA.76.042319} {\bibfield  {journal} {\bibinfo  {journal} {Phys.
  Rev. A}\ }\textbf {\bibinfo {volume} {76}},\ \bibinfo {pages} {042319}
  (\bibinfo {year} {2007})}\BibitemShut {NoStop}%
\bibitem [{\citenamefont {Kapit}(2015)}]{Kapit2015}%
  \BibitemOpen
  \bibfield  {author} {\bibinfo {author} {\bibfnamefont {E.}~\bibnamefont
  {Kapit}},\ }\href {\doibase 10.1103/PhysRevA.92.012302} {\bibfield  {journal}
  {\bibinfo  {journal} {Phys. Rev. A}\ }\textbf {\bibinfo {volume} {92}},\
  \bibinfo {pages} {012302} (\bibinfo {year} {2015})}\BibitemShut {NoStop}%
\bibitem [{\citenamefont {Gardiner}\ and\ \citenamefont
  {Collett}(1985)}]{Gardiner1985}%
  \BibitemOpen
  \bibfield  {author} {\bibinfo {author} {\bibfnamefont {C.~W.}\ \bibnamefont
  {Gardiner}}\ and\ \bibinfo {author} {\bibfnamefont {M.~J.}\ \bibnamefont
  {Collett}},\ }\href {\doibase 10.1103/PhysRevA.31.3761} {\bibfield  {journal}
  {\bibinfo  {journal} {Phys. Rev. A}\ }\textbf {\bibinfo {volume} {31}},\
  \bibinfo {pages} {3761} (\bibinfo {year} {1985})}\BibitemShut {NoStop}%
\bibitem [{\citenamefont {Mezzacapo}\ \emph {et~al.}(2014)\citenamefont
  {Mezzacapo}, \citenamefont {Lamata}, \citenamefont {Filipp},\ and\
  \citenamefont {Solano}}]{Mezzacapo2014}%
  \BibitemOpen
  \bibfield  {author} {\bibinfo {author} {\bibfnamefont {A.}~\bibnamefont
  {Mezzacapo}}, \bibinfo {author} {\bibfnamefont {L.}~\bibnamefont {Lamata}},
  \bibinfo {author} {\bibfnamefont {S.}~\bibnamefont {Filipp}}, \ and\ \bibinfo
  {author} {\bibfnamefont {E.}~\bibnamefont {Solano}},\ }\href {\doibase
  10.1103/PhysRevLett.113.050501} {\bibfield  {journal} {\bibinfo  {journal}
  {Phys. Rev. Lett.}\ }\textbf {\bibinfo {volume} {113}},\ \bibinfo {pages}
  {050501} (\bibinfo {year} {2014})}\BibitemShut {NoStop}%
\bibitem [{\citenamefont {Lechner}\ \emph {et~al.}(2015)\citenamefont
  {Lechner}, \citenamefont {Hauke},\ and\ \citenamefont
  {Zoller}}]{Lechner2015}%
  \BibitemOpen
  \bibfield  {author} {\bibinfo {author} {\bibfnamefont {W.}~\bibnamefont
  {Lechner}}, \bibinfo {author} {\bibfnamefont {P.}~\bibnamefont {Hauke}}, \
  and\ \bibinfo {author} {\bibfnamefont {P.}~\bibnamefont {Zoller}},\ }\href
  {\doibase 10.1126/sciadv.1500838} {\bibfield  {journal} {\bibinfo  {journal}
  {Science Advances}\ }\textbf {\bibinfo {volume} {1}} (\bibinfo {year}
  {2015}),\ 10.1126/sciadv.1500838},\ \Eprint
  {http://arxiv.org/abs/http://advances.sciencemag.org/content/1/9/e1500838.full.pdf}
  {http://advances.sciencemag.org/content/1/9/e1500838.full.pdf} \BibitemShut
  {NoStop}%
\bibitem [{\citenamefont {Zunkovic}\ and\ \citenamefont
  {Prosen}(2010)}]{Zunkovic2010}%
  \BibitemOpen
  \bibfield  {author} {\bibinfo {author} {\bibfnamefont {B.}~\bibnamefont
  {Zunkovic}}\ and\ \bibinfo {author} {\bibfnamefont {T.}~\bibnamefont
  {Prosen}},\ }\href {http://stacks.iop.org/1742-5468/2010/i=08/a=P08016}
  {\bibfield  {journal} {\bibinfo  {journal} {Journal of Statistical Mechanics:
  Theory and Experiment}\ }\textbf {\bibinfo {volume} {2010}},\ \bibinfo
  {pages} {P08016} (\bibinfo {year} {2010})}\BibitemShut {NoStop}%
\end{thebibliography}%

\end{document}